\documentclass[final,authoryear,1p,times]{elsarticle}

\usepackage{graphicx}

\usepackage{amssymb}
\usepackage{amsmath}
\usepackage{lineno}
\usepackage{numcompress}
\usepackage[T1]{fontenc}
\usepackage[latin1]{inputenc}
\usepackage{color}
\usepackage{pgf}

\journal{J. Atm. Sol-Ter. Phys.,}

\begin{document}

\begin{frontmatter}



\title{Tsunami effects on the Z component of the geomagnetic field}


\author[label1,label2]{Klausner V.}
\author[label4]{Domingues M. O.}
\author[label2]{Mendes O.}
\author[label1,label3]{Papa A. R. R.}

\address[label1]{National Observatory - ON 20921-400, RJ, Brazil}
\address[label2]{DGE/CEA/National Institute for Space Research - INPE 12227-010 S\~ao Jos\'e dos Campos, SP, Brazil}
\address[label3]{Rio de Janeiro State University - UERJ, RJ, Brazil}
\address[label4]{LAC/CTE/National Institute for Space Research - INPE 12227-010 S\~ao Jos\'e dos Campos, SP, Brazil}

\begin{abstract}
The vertical component (Z) of the geomagnetic field observed by ground-based observatories of the INTERMAGNET network has been used to analyze the effects of the movement of electrically conducting sea water through the geomagnetic field due to a propagation of a tsumani.
The purpose of this work is to study the geomagnetic variations induced by the tsunamis occurred at 26 December, 2004, 27 February, 2010 and 11 March, 2011.
For each case study, we selected four magnetic stations belonging to the INTERMAGNET programme that were influenced or more direct affected by the tsumani.
To detect these disturbances in the geomagnetic data,  the discrete wavelet technique have been used in four levels of decomposition.
We were able to detect the localized behavior of the geomagnetic variations induced by the movement of electrically conducting sea-water through the geomagnetic field, \textit{i.\,e.}, the identification of transients related to the tsunamis.
As well, using the minutely magnetogram data, it was able to localize the initial phase and time of the tsunami maximum height.
The first interpretation of the results suggests that discrete wavelet transform can be used to characterize the tsumani effects on the geomagnetic field, but need further study.
\end{abstract}


\end{frontmatter}
\linenumbers


\section{Introduction}
\label{Introduction}
In the treatise ``De Magnete'' as early as 1600, the english researcher Gilbert showed the predominately dipolar character of the terrestrial magnetic field. 
Next in the early nineteen century, other english researcher Gauss improved the magnetic field observation techniques and introduced the spherical harmonic method for geomagnetic field analysis. 
Based on a called ``Spherical Harmonic Analysis'', a mathematical method was developed to separate the external and the internal contributions to the surface field by a unique global analysis of the Earth's main field  \citep[see][and references therein]{Campbell1997}.

The geomagnetic field is described as a complicated function of space and time. Ground based magnetic measurements show a repetitive diurnal variation on geomagnetically quiet days. But there is a great variety of irregular variations that occur from time to time, the ``disturbance fields''. 
Periods of great disturbance are called, by analogy with the weather, ``magnetic storms'' \citep{Parkinson1983}.

Some evidence of the influence of oceanic tides on the magnetic daily variation has been obtained by \Citet{LarsenCox1966}. 
They found small semidiurnal variations of the Z component at a coastal site (Cambria, California) and at two island stations (Honolulu and San Miguel) that could not be explained by the atmospheric tidal theory.
They suggested that these variations must be due predominantly to oceanic tides. 
It is important to mention here that the conductivity of the ocean does not vary significantly with time, unlike the ionospheric conductivity.
As a consequence, the seasonal variation of the oceanic contribution is expected to be smaller than the ionospheric contribution \citep{Cuetoetall2003}.

\Citet{Manoj2011} observed the geomagnetic contributions due to the moderate tsumani in the Pacific ocean provoked by the Chilean earthquake of $8.8$ in magnitude at three different magnetic stations (PPT, HUA and IPM).
Their observations presented a variation of $1~\mathrm{nT}$ in the vertical component of the magnetic field (Z) during the time of the tsumani effects.
Movement of electrically conducting sea-water through the geomagnetic field generates an electromotive force that induces electric fields, currents and secondary magnetic fields.
In other words, tsunamis can produce perturbations in the Earth's magnetic field by electro-magnetic induction \citep[see][and references therein]{Manoj2011}.

In this work, we focused on the survey of geomagnetic variations induced by some selected tsunamis: one of 26 December 2004, other of 27 February 2010 and another of 11 March, 2011.
As the wavelet transforms had turn out to be a very useful tool in atmospheric signal analysis \citep[see][and references therein]{Domingues2005}, the discrete wavelet technique have been selected and used with four levels of decomposition in order to detect the small singularities in the geomagnetic data.
For a good revision on the application of discrete wavelet on Space Weather Research, \Citet{Domingues2005}, \Citet{MendesMag2005} and \cite{MendesdaCostaetal:2011} can be consulted.
Thus, this work aims to verify the use of the wavelet technique as a way to identify the effects related to tsunamis on the geomagnetic field components, particularly the Z-component.
The contents are composed by: in Section~\ref{Geomagnetic disturbance mechanism}, the presentation of the geomagnetic disturbance mechanism related to the tsunami effects;
in Section~\ref{Discrete Wavelet Transform (DWT) methodology}, applied the methodology;
in Section~\ref{Magnetic Data}, the data;
in Section~\ref{Results and analyses}, the results achieved by the analysis; and 
in Section~\ref{Conclusions}, the conclusions of our work.

\section{Geomagnetic disturbance mechanism}
\label{Geomagnetic disturbance mechanism}

Faraday originally predicted that ocean waves could induce electrical fields, which are also called by ''motional induction``.
The induction mechanism process consists of the flow of the ocean water, as an electrically conducting fluid, through the Earth's main magnetic field.
The salts dissolved in the sea water are made of ions that can be deflected by the Lorentz force which acts in a perpendicular direction both to the velocity and magnetic field vectors \citep{Tyleretal:2003, Manoj2006}.
As consequence of the motion of sea water with velocity $\overrightarrow{V}$ across the geomagnetic field $\overrightarrow{B}$, a conduction current $\overrightarrow{J}$ is produced expressed by:
\begin{equation}
 \overrightarrow{J}=\sigma \; (\overrightarrow{E}+\overrightarrow{V} \times \overrightarrow{B}) ,
\end{equation}
\noindent
where $\sigma$ is the electrical conductivity, $\overrightarrow{E}$ is the stationary electric field and $\overrightarrow{V} \times \overrightarrow{B}$ is the induced electric field.
Discarding any displacement current, the Ampere's law is used to determine an induced magnetic field  $\overrightarrow{b}$ given by:
\begin{equation}
 \overrightarrow{\nabla} \times \overrightarrow{b}=\mu_o\overrightarrow{J},
\end{equation}
\noindent
where $\mu_o=4\pi10^{-7} \;\; henry/m$. 
It is known that electric and magnetic fields are related by Faraday's law given by: 
\begin{equation}
 \overrightarrow{\nabla} \times \overrightarrow{E}=\dfrac{\partial \overrightarrow{b}}{\partial t}.
\end{equation}

Applying the same mathematical development used by \Citet{Podney1975}, the velocity field present in the problem can be described as a curl of a vector stream function $\overrightarrow{\Psi}$ (the stream understood as the flow of the ocean water), \textit{i.e.}:
\begin{equation}
 \overrightarrow{V}= \overrightarrow{\nabla} \times \overrightarrow{\Psi}.
\end{equation}
In analogy, the induced magnetic field can be expressed by a vector potential $\overrightarrow{A}$, given by:
\begin{equation}
 \overrightarrow{b}= \overrightarrow{\nabla} \times \overrightarrow{A}.
\end{equation}
Thus the relation between the vector stream function $\overrightarrow{\Psi}$  and the geomagnetic field $\overrightarrow{B}$ as a function of the vector potential $\overrightarrow{A}$  can be established as:
\begin{equation}
\nabla^{2} \; \overrightarrow{A} - \mu_o \; \sigma \; \dfrac{\partial \overrightarrow{A}}{\partial t}= -\mu_o \; \sigma \; \overrightarrow{\nabla} \times (\overrightarrow{\Psi} \times \overrightarrow{B}) .
\end{equation}

In a short description, the induced magnetic field generated by the ocean can be classified by two components: toroidal and poloidal.
The toroidal component is generated by the electrical currents closing in the vertical plane and can reach up to $100~\mathrm{nT}$ but is confined to the ocean and the upper crust.
The poloidal component is much weaker, between $1$ and $10~\mathrm{nT}$, and arises from the electrical currents closing in the horizontal plane, however it can reaches outside of the ocean to remote lands and satellite locations \citep[see][and references therein]{Tyleretal:2003, Manoj2006}.

The sea water is subject to two kinds of geomagnetic fluctuations: periodic (lunar and solar variations) and transient disturbances (geomagnetic storms), as discussed by \Citet{Longuet1954}.
Also, the ocean flow can be occasionally disturbed by tsunamis that can increase the ocean wave speed for few centimeters per second and that affects the entire water column. Consequently, as demonstrated before, to increase the ocean flow will also increase the magnetic induced field.

Considering quiet conditions, \Citet{Tyleretal:1999} discussed that the major difficulty to determine the magnetic ocean generated signals are the weak values compared to the signals from other sources.
This is the reason to take advantage of some analysis features of the wavelet technique.

\section{Discrete Wavelet Transform (DWT) methodology}
\label{Discrete Wavelet Transform (DWT) methodology}

The wavelet transforms were better and broadly formalized thanks to mathematicians, physicist, and engineers efforts \citep[e.g.,][]{Morlet1983}. 
In the domain of geophysics applications, the main characteristic of the wavelet technique is the introduction of time-frequency decomposition \citep[e.g.,][]{Domingues2005, MendesMag2005, MendesdaCostaetal:2011}. 
In the 1990s several important ideas and applications concerning wavelet were developed \citep[e.g.,][]{Daubechies1992, Chui1992Int, Chui1992Wa, Chui1994}.

The wavelet analysis could show that the larger amplitudes of the wavelet coefficients are locally associated with abrupt signal changes. 
In recent work \citep{Domingues2005, MendesMag2005, MendesdaCostaetal:2011}, a method for the detection of the transition region and the exact location of this discontinuities due to geomagnetic storms was implemented.

As a basis for this work, this mathematical method is presented in an abbreviated way. 
The multiresolution analysis (AMR) is a mathematical tool used to build up wavelet functions \citep[e.g.,][]{Daubechies1992, Mallat1991}.
Then it is possible to build up wavelet functions using a pair of vectors $\{V^j, \phi^j\}$, in such a way that there are a sequence of embedded approximating spaces and the functions $V^j\subset V^{j+1}$ and $\phi_k^j$ form a Riesz basis for $V^j$, being $V^j=span\{\phi_k^j(t)\}$.
In this technique, a mother-wavelet is generated from a scaling function. 
It obeys the following scale relation \citep[more details in][]{Mallat1991}:
    \begin{equation}
          \phi(t)=2 \sum\limits_{k\in Z}h(k)\phi(2t-k) \:\: \mbox{ (Scale relation)}
    \end{equation}
and $h(k)$ is the low pass-filter, also called scale filter coefficients.
The family of scale relations are represented by:
    \begin{equation}
          \phi_k^j(t)= 2^{j/2}\phi(2^jt-k),\qquad j,k\in Z
    \end{equation}
They are called scale functions and in frequency domain are given by:
    \begin{equation}
          \hat\phi(\xi)=H(\xi/2)\hat\phi(\xi/2)
    \end{equation}
where $H(\xi)=\sum\limits_{k\in \ h(k)}e^{-\imath k \xi}$ is a low pass filter associate to $\phi$.
The following relation holds:
    \begin{equation}
          V^{j+1}=V^j\oplus W^j
    \end{equation}
The spans $W^j$ have the difference of information between $V^j$ and $V^{j+1}$.
If the $\psi$ function form a Riesz base of $W^j$, it is called wavelet function: 
 \begin{equation}
\psi_k^j(t)=2^{j/2}\psi(2^jt-k)\qquad
\psi(t)=2\sum\limits_{k \in Z}g(k)\psi(2t-k)
\end{equation}
where $g(k)=(-1)^{k+1}h(-k+1)$ is a high pass filter.

The wavelet and scale functions satisfy the orthogonality condition, $\langle\phi_k^j, \psi_l^j\rangle=0$, $\langle\psi_k^j,\psi_l^m\rangle = \delta_{j,m}\delta_{k,l}$.
The AMR tool is useful to study the function in $L^2(\Re)$.
The difference of information between $V^j$ and $V^{j+1}$ is given by $W^j$, \textit{i.\,e.},
\begin{equation}
 \left(\Pi^{j+1}-\Pi^j\right)f(t)=Q^jf(t),
\end{equation}
where the projections in $V^j$ and $W^j$ are, respectively:
   \begin{equation}
       \Pi^jf(t)=\sum\limits_k\langle f,\phi_k^j\rangle\phi_k^j(t) \qquad  
       Q^jf(t)=\sum\limits_k\langle f,\psi_k^j\rangle\psi_k^j(t),
   \end{equation}
 We obtain in multi-level $j_0 < j$ the coefficients expression:
   \begin{equation}\label{eq:coefW}
    c_k^j=\langle f,\phi_k^j\rangle \qquad  d_k^j=\langle f,\psi_k^j\rangle
   \end{equation}
We can obtain with a change of base $\{\phi_k^{j+1}\}\leftrightarrow\{\phi_k^{j_0}\}\cup \{\psi_k^{j_0}\}\ldots\cup \{\psi_k^j\} $, the equation:
    \begin{equation}
     \sum\limits_k c_k^{j+1}\phi_k^{j+1}(t)= \sum\limits_k c_k^{j_0}\phi_k^{j_0}(t)+\sum\limits_{m=j_0}^{j} \sum\limits_k d_k^{m}\psi_k^{m}(t)
    \end{equation}
 In the DWT the equation (\ref{eq:coefW}) is manipulated jointly with to the scale relations:
  \begin{equation} \label{eq:coeffa}
       c_k^j=\sqrt 2 \sum\limits_m h(m-2k)c_m^{j+1} \qquad
       d_k^j=\sqrt 2 \sum\limits_m g(m-2k)c_m^{j+1} 
   \end{equation}
   In this work we use the Daubechies (db1) wavelet function of order 1 (Haar wavelet). In this case the coefficients $h = [\frac{\sqrt{2}}{2},\frac{\sqrt{2}}{2}]$ and consequently, $g = [\frac{\sqrt{2}}{2}, -\frac{\sqrt{2}}{2}]$.

The wavelet transform can be used in the analysis of non-stationary signals to rescue  information on the frequency or scale variations of those signals and to detect the localization of theirs structures in time and/or in space.
Due to the double localization property of the wavelet function, the wavelet transform is said to be of local type in time-frequency, with time and frequency resolutions inversely proportional. 
As discussed by \cite{Domingues2005}, one consequence of the properties of the wavelet analysis is the possibility of showing that the wavelet coefficients amplitude is associated with abrupt signal variations or "details" of higher frequency \citep[see also][]{Meyer:1990,Chui1992Wa}. 
Selecting a wavelet function which closely matches the signal to be processed is of utmost importance in wavelet applications. 
A tutorial on the main properties and applications of the wavelet analysis can be found, for instance, in \Citet{Domingues2005} and \Citet{MendesMag2005}.

A hard thresholding process has been applied to the wavelet coefficients for the first three decomposition levels to identify the storm time in the magnetograms, as has been discussed by \cite{MendesMag2005}. 
Thresholding concerns the process of setting to zero certain coefficients in an effort to highlight significant information, in this case the shock-like transient phenomena that appear in the main phase of the magnetic storms \citep[see][]{Domingues2005,MendesMag2005}.
It was found that $2^{j-1}$ could be used as threshold sets for the decomposition levels $(d^j)^2$, because according to Meyer theorem, the multilevel threshold sets may have a multiplicity of $2$ \citep{Meyer:1990}. 

In this analysis, we choose to use the Haar wavelet and the sampling rate of $1$ min, as consequence of the time resolution of the magnetic data, with the pseudo-periods of the first four levels 2, 4, 8 and 16 min. 
The Haar wavelet is considered the most simple orthogonal analyzing wavelet function.
However the discrete wavelet transform (DWT) using Haar wavelet can detect abrupt variations, \textit{i.\,e.}, one localization feature in the physical space.
The Haar wavelet has a compact support in physical space and a large support in Fourier space.
In other words, Haar wavelet has a better localization on time than on frequency, what is expected by the Heisenberg's uncertainty principle \citep[as discussed in][]{Daubechies1992}.

\section{Magnetic Data}
\label{Magnetic Data}

In this section, we first describe the data used to study the geomagnetic variations due to the tsunami-generated magnetic fields.
The three tsunami's events are presented.
For each event, we have chosen four ground magnetic measurements.
We have also selected the tide-gauge measurements at or nearby the chosen magnetic stations, only for guiding purposes.

We selected magnetic stations belonging to the INTERMAGNET programme (http://www. intermagnet.org) that were influenced or more direct affected by the tsumanis.
By international agreement, there are two systems that can represent the Earth's magnetic field: the XYZ and the HDZ system \citep[see][and references therein]{Campbell1997}.
The X, Y and Z stand for northward, eastward and vertical into the Earth directions, the H, D and Z stand for horizontal component, declination (angular direction of the horizontal component) and vertical (into the Earth).
The H-component is more affected by the solar-magnetospheric interactions, consequently, also the X- and Y-component.
These variations, specially those associated to the ring current, have a major contribution on the magnetograms at stations located at low and mid-latitude regions.
Due to the Z-component be less affected than the H-component, we decided to use the Z-component to detect the geomagnetic variations induced by the tsunami.
Using a different technique in a work with similar purpose, \Citet{Manoj2011} also used the Z-component to observe the variations in the Z-component provoked by the Chilean tsumani, 27 February, 2010.

The intensity of the geomagnetic disturbance in each day is described by indices.
There are different indices that can be used depending on the character and the latitude influences in focus.
Kp, AE and Dst and their derivations are the most used geomagnetic indices.
The Kp index is a number from 0 to 9 obtained as the mean value of the disturbance levels within 3-h interval observed at 13 subauroral magnetic stations \citep[see][]{Bartels1957}.
The minutely AE index (sometimes $2.5$ minute interval) is obtained by the H-component measured from the magnetic stations located at auroral zones.
The index most used in low and mid-latitudes is the Dst index.
It represents the variations of the H-component due to changes of the ring current.
The characteristic signature of a magnetic storm is a depression in the horizontal component of the Earth's magnetic field H and this depression is shown by the Dst index. 

In this work, we analyzed the SYM-H index to identify periods of disturbed geomagnetic field conditions.
The SYM-H is essentially the same as the traditional hourly Dst index.
The main characteristic of the $1$ minute time resolution SYM-H index is that the solar wind dynamic pressure variation are more clearly seen than indices with lower time resolution.
Its calculation is based on magnetic data provided by 11 stations of low and medium latitude.
Only six of the stations are used for its calculation of each month, some stations can be replaced by others depending on the data conditions.

We focused on the survey of geomagnetic variations induced by the tsunamis occurred at 26 December, 2004, 27 February, 2010 and 11 March, 2011.
Related to the magnetic disturbance level, these events of tsunami occurrence happened during quiet time (27 February, 2010) and disturbed geomagnetic conditions (weak - 26 December, 2004, and moderate storms, 11 March, 2011).

On 26 December 2004 the Sumatra-Andaman earthquake occurred with $9.1 M_w$ ($M_w$ means moment magnitude of the Richter scale) at $00:59$ UT (Universal Time).
The epicenter coordinates are Lat. $3.4^{\circ}$ and Long. $95.7^{\circ}$ at $30~\mathrm{km}$ depth.
\cite{Tsujietal:2006} carried out tsunami height measurements on the Andaman seacoast.
They found the maximum tsunami height was $19.6$ m at Ban Thung Dap (Lat. $9.03^{\circ}$ and Long. $98.26^{\circ}$) and the second highest was $15.8$ m at Ban Nam Kim (Lat.$8.86^{\circ}$ and Long.$98.28^{\circ}$).
The tsunami not only propagated throughtout the Indian ocean but also into the Pacific and Atlantic oceans.
\cite{Murtyetal:2005} discussed through a simple analytical model that the tsunami flux from the Indian ocean into the Pacific ocean is greater than the flux into the Atlantic ocean.
This leakage of tsunami energy into the Pacific and Atlantic oceans occurs through the gaps between Australia and Antarctica and between Africa and Antarctica, respectively.
The maximum tsunami amplitudes were of $0.65~\mathrm{m}$ on the Pacific coast and $0.3~\mathrm{m}$ on the Atlantic coast as discussed by \Citet[][and references within]{Murtyetal:2005}.

\begin{figure}[htb]
	\centering
		\includegraphics[width=10cm]{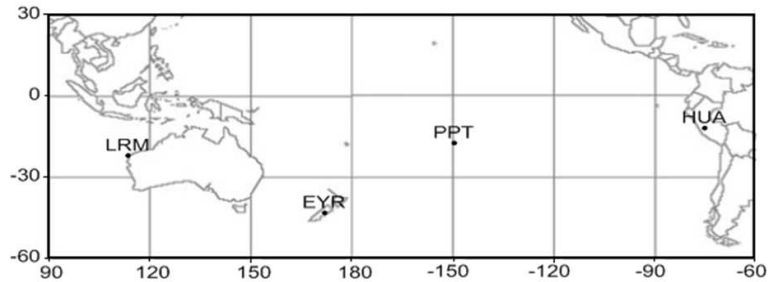}
	\label{fig:MapThaiStations}
	\caption{Geographical localization of the stations used in the study of the Sumatra-Andaman tsunami 2004}
\end{figure}

To study the geomagnetic variations due to the Sumatra-Andaman tsunami, the magnetic stations considered were: Eyrewell (EYR), Huancayo (HUA), Learmonth (LRM) and Papeete (PPT), with the geographic and geomagnetic coordinates presented in Table~\ref{table:ABBcodeThai}.
Figure~\ref{fig:MapThaiStations} display the magnetic stations distribution and their IAGA codes.
The selected tide-gauge measurements, only for guiding purposes, are located at Cocos Island (Lat. $-12.12^{\circ}$, Long. $96.90^{\circ}$), Jackson Bay (Lat. $-43.97^{\circ}$, Long. $168.62^{\circ}$), Nuku Hiva (Lat. $-08.82^{\circ}$, Long. $140.20^{\circ}$) and Callao (Lat. $-12.08^{\circ}$, Long. $-77.13^{\circ}$).

\begin{table}[ht]
 \caption{INTERMAGNET network of geomagnetic stations for the study of the Thai tsunami.}
\centering
\begin{tabular}{c c c c c }
\hline
Station & \multicolumn{2}{c}{Geografic coord.} & \multicolumn{2}{c}{Geomagnetic coord.}\\
\cline{2-5}
          & Lat.($\,^{\circ}$) & Long.($\,^{\circ}$) & Lat.($\,^{\circ}$) & Long.($\,^{\circ}$)  \\[0.5ex]
\hline
\\
EYR    &-43.41     &172.35         &-46.92   &-106.22\\
HUA    &-12.04     &-75.32         &-1.73   &-3.44\\
LRM   &-22.22      &114.10         &-32.28    &-173.53\\
PPT    &-17.75     &-149.57         &-15.20    &-74.83\\[1ex]
\hline
\end{tabular}
\\Source: http://wdc.kugi.kyoto-u.ac.jp/igrf/gggm/index.html (2005)
\label{table:ABBcodeThai}
\end{table}

Near coast of the central Chile, on the $27$~th of February, 2010 at  $06:34$ UT occurred an earthquake with magnitude $8.8\;M_w$.
The epicenter was located on Lat. $-36.1^{\circ}$ and Long. $-72.6^{\circ}$ at $55~\mathrm{km}$ depth.
As reported by \cite{Pararas-Carayannis2010}, shortly after the earthquake, tsunami waves hit the coastal area of the Central Chile.
The tsunami overtook the coastal cities as Talcahuano, Coquimbo, Antofasta and Caldera, as well as the Juan Fern\'andez Islands.
The NOAA Pacific Warning Center released a bulletin number $018$ and a tsunami warning was issued at $00:12$ UT on 28 February, 2010 for Chile, Peru, Ecuador, Colombia, Antarctica, Panama, Costa Rica, Nicaragua, Pitcairn, Honduras, El Salvador, Guatemala, French Polynesia, Mexico, Cook Islands, Kiribati, Kermadec Island, Niue, New Zealand, Tonga, American Samoa, Samoa, Jarvis Island, Wallis-Futuna, Tokelau, Fiji, Australia,   Hawaii, Palmyra Island, Johnston Island, Marshall Island, Midway Island, Wake Island, Tuvalu, Vanuatu, Howland-Baker, New Caledonea, Solomon Island, Nauru, Kosrae, Papaua New Guinea, Pohnpei, Chuuk, Marcus Island, Indonesia, North Marianas, Guam, Yap, Belau, Philippines and Chinese Taipei.

\begin{figure}[ht]
	\centering
		\includegraphics[width=8cm]{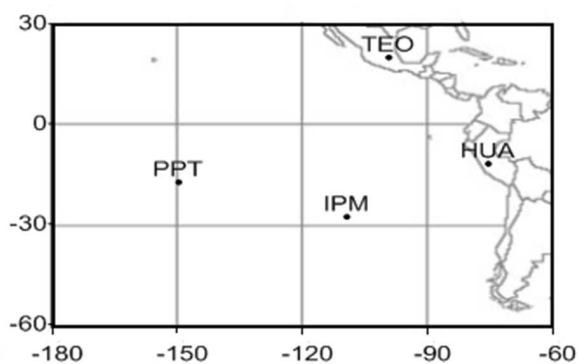}
	\label{fig:MapChileStations}
	\caption{Geographical localization of the stations used in the study of the Chilean tsunami, 2010.}
\end{figure}

The magnetic stations considered for this event were: Huancayo (HUA), Easter Island (IPM), Papeete (PPT) and Teoloyucan (TEO), with the geographic and geomagnetic coordinates presented in Table~\ref{table:ABBcodeChile}.
Figure~\ref{fig:MapChileStations} display the magnetic stations distribution and their IAGA codes.
We selected the same three stations used by \Citet{Manoj2011} to study the geomagnetic contributions due to the Chilean tsumani and included the station of TEO located in Mexico.
In their study only the IPM station showed a periodic variation of $1~\mathrm{nT}$ in the vertical component (Z) caused by the tsunami started at $11:35$ UT, and the other two stations did not show concurrent variations.

The selected tide-gauge measurements are located at Callao La Punta (Lat. $-12.01^{\circ}$ and Long. $-77.17^{\circ}$), Papeete (Lat. $-17.75^{\circ}$ and Long. $-149.57^{\circ}$) and Acapulco (Lat. $16.83^{\circ}$ and Long. $-99.92^{\circ}$).

\begin{table}[ht]
 \caption{INTERMAGNET network of geomagnetic stations for the study of the Chilean tsunami.}
\centering
\begin{tabular}{c c c c c }
\hline
Station & \multicolumn{2}{c}{Geografic coord.} & \multicolumn{2}{c}{Geomagnetic coord.}\\
\cline{2-5}
          & Lat.($\,^{\circ}$) & Long.($\,^{\circ}$) & Lat.($\,^{\circ}$) & Long.($\,^{\circ}$)  \\[0.5ex]
\hline
\\
HUA    &-12.04     &-75.32         &-1.99   &-3.03\\
IPM    &-27.90      &-109.25         &-19.63    &-34.47\\
PPT    &-17.75     &-149.57         &-15.20    &-74.49\\
TEO    &19.75     &-99.19         &28.45    &-29.07\\ [1ex]
\hline
\end{tabular}
\\Source: http://wdc.kugi.kyoto-u.ac.jp/igrf/gggm/index.html (2010)
\label{table:ABBcodeChile}
\end{table}

On 11 March, 2011, an earthquake occurred in the Japanese coast at $05:46$ UT with $8.9\;M_w$.
The epicenter located at Lat. $38.3\,^{\circ}$ and Long. $142.4\,^{\circ}$, near to the east coast of Honshu, at $24~\mathrm{km}$ depth.
That earthquake induced a ``major'' tsunami that affected all the Japanese coast and transversed the Pacific Ocean eastward.

\begin{figure}[htb]
	\centering
		\includegraphics[width=8cm]{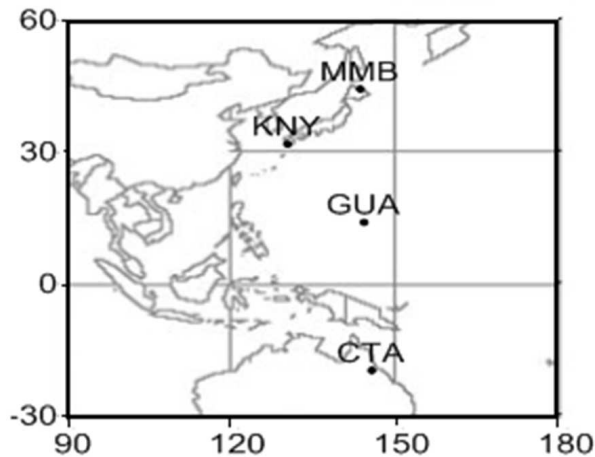}
	\label{fig:MapJapStations}
	\caption{Geographical localization of the stations used in the study of the Japanese tsunami 2011}
\end{figure}

As we have done before , we also choose four stations that belong to the INTERMAGNET programme that were influenced or more directly affected by the Japanese tsunami of March 11, 2011 to study the geomagnetic variations.
The stations considered in this analysis were: Charters Towers (CTA), Guam (GUA), Kanoya (KNY) and Memambetsu (MMB), with
geographic and geomagnetic coordinates presented in Table~\ref{table:ABBcodeJap}.
Figure~\ref{fig:MapJapStations} display the magnetic stations distribution and their IAGA codes.

The selected tide-gauge measurements are located at Hanasaki (Lat. $43.28^{\circ}$ and Long. $145.57^{\circ}$), Tosashimizu (Lat. $32.78^{\circ}$ and Long. $132.92^{\circ}$), Pago Bay (Lat. $13.43^{\circ}$ and Long. $144.80^{\circ}$) and Cape Ferguson (Lat.$-19.28^{\circ}$ and Long.$147.06^{\circ}$).

\begin{table}[ht]
 \caption{INTERMAGNET network of geomagnetic stations for the study of the Japanese tsunami.}
\centering
\begin{tabular}{c c c c c }
\hline
Station & \multicolumn{2}{c}{Geografic coord.} & \multicolumn{2}{c}{Geomagnetic coord.}\\
\cline{2-5}
          & Lat.($\,^{\circ}$) & Long.($\,^{\circ}$) & Lat.($\,^{\circ}$) & Long.($\,^{\circ}$)  \\[0.5ex]
\hline
\\
CTA    &-20.10     &146.30         &-27.59   &-138.58\\
GUA    &13.28      &144.45         &05.18    &-144.20\\
KNY     &31.42     &130.88         &22.06    &-158.71\\
MMB     &35.44     &144.19         &27.09    &-146.95\\ [1ex]
\hline
\end{tabular}
\\Source: http://wdc.kugi.kyoto-u.ac.jp/igrf/gggm/index.html (2010)
\label{table:ABBcodeJap}
\end{table}

\section{Results and analysis}
\label{Results and analyses}

Initially the wavelet treatment is done on the SYM-H Dataset, which is a processed data.
According to the theory used, expected results are obtained. 
After that the same treatment is done directly on the magnetometer dataset for each tsunami event. 
The graphical results are presented and discussed validating the identifications of the tsunami occurrences (Dec. 26, 2004, Feb. 27, 2010, and Mar. 11, 2011).

From theory the wavelet coefficients are supposed to allow identifying the induced magnetic variations due to tsunamis, because the technique can detect physical discontinuities in the vertical component of the geomagnetic field. 
When the magnetosphere is under quiet conditions the behavior of the recorded Z-component should be much smoother than its behavior in the disturbed periods, making easy the identification of the variations only induced by the propagation of the tsunami.
Using a very localized wavelet, we will be able to detect these small geomagnetic variations. 
The Haar wavelet present a short compact support that provides a better local characterization of the signal and it may be the most convenient to represent times series with abrupt variations or steps.

\subsection{Analysis on SYM-H Dataset}

\begin{figure}[hbt]
\noindent
\centering
\begin{tabular}{cc}
(a) $26^{th}$ December, $2004$  & (b) $27^{th}$ December, $2004$ \\
\includegraphics[height=6.5cm,width=6.5cm]{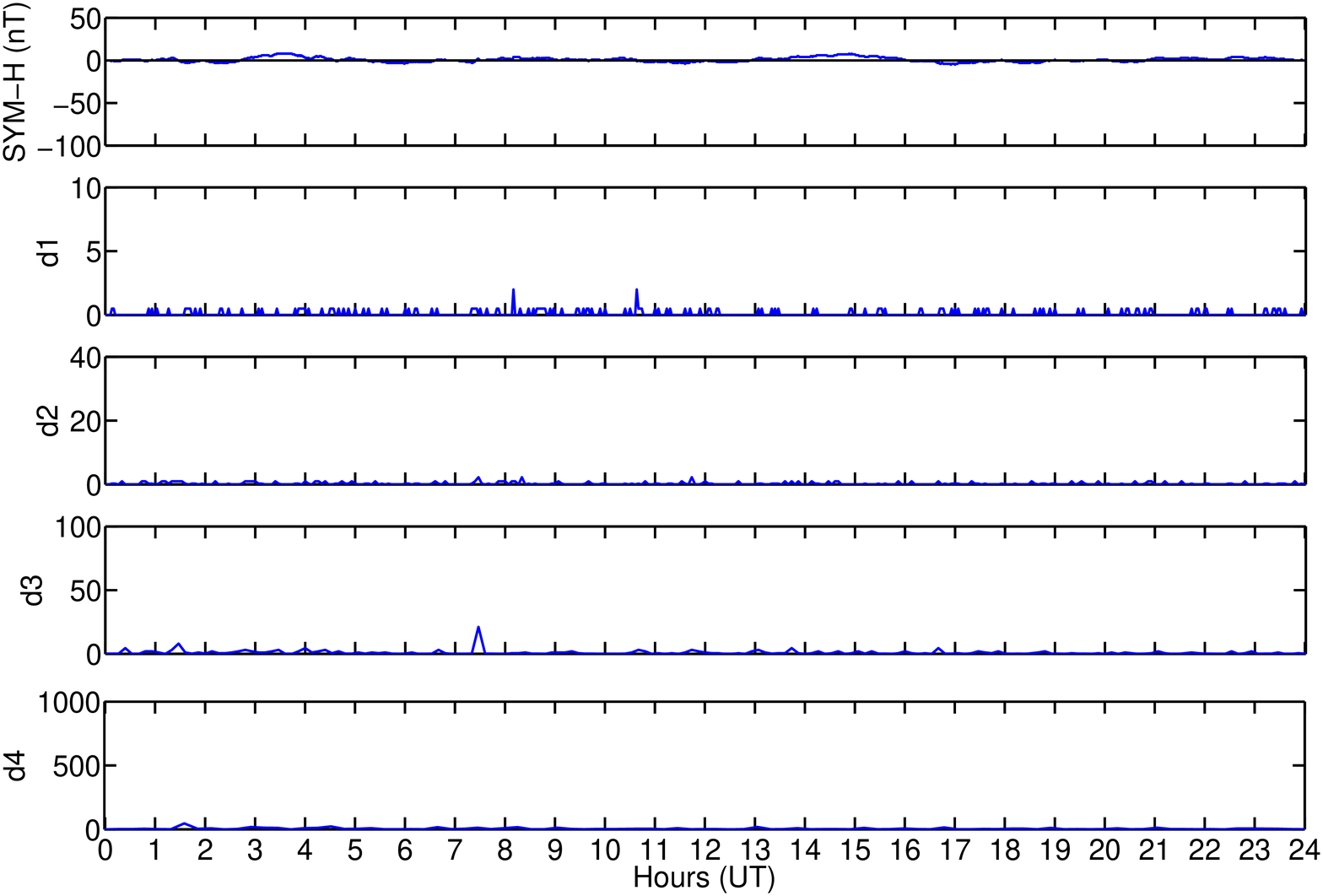}&
\includegraphics[height=6.5cm,width=6.5cm]{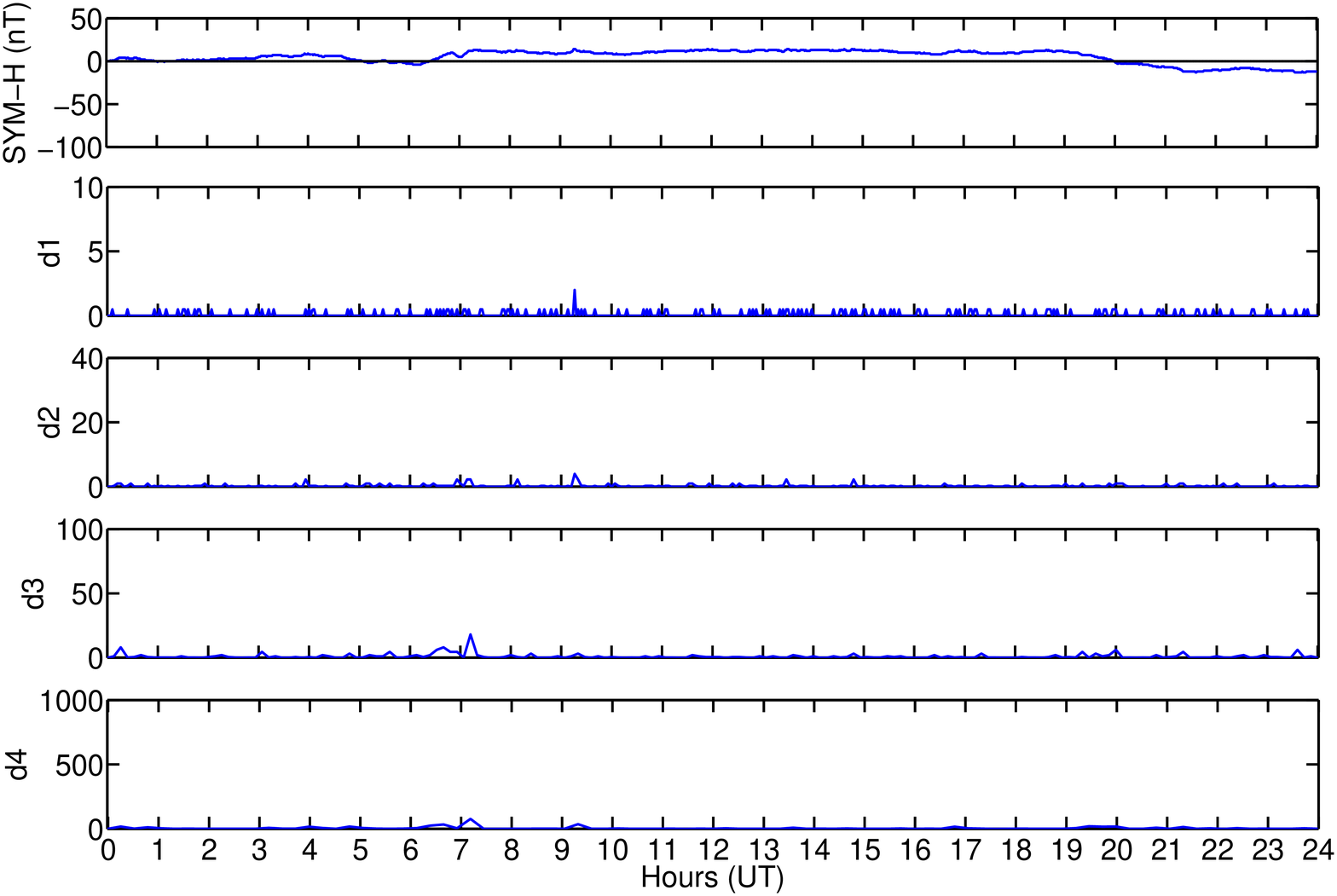}\\
\end{tabular}
\caption{Graphics of SYM-H with the corresponding wavelet coefficients $(d^j)^2$ for $j=1,2,3,4$. 
The panel corresponds to a tsumani occurrence, (a) the arrival day and (b) the day after the Sumatra-Andaman tsunami 2004.}
\label{fig:SYM-H}
\end{figure}

Figure~\ref{fig:SYM-H} presents the graphical results of wavelet analysis for sample periods related to tsunami occurrences.
In each part, at the top there is the SYM-H record and at the bottom respectively from the first to the fourth level of the wavelet decomposition. 
Based on the theory, the amplitude of the wavelet coefficients should identify the transient behavior in the dataset.

In the figure it is shown the discrete wavelet decomposition applied to minutely SYM-H index using Daubechies orthogonal wavelet family $1$.
From top to bottom each panel shows, the SYM-H and the first four levels of wavelet decomposition ($(d^j)^2$ where $j=1,2,3,4$) for the day corresponding to the tsumani of 2004 due to the Sumatra-Andaman earthquake (Figure~\ref{fig:SYM-H}a) and the day after (Figure~\ref{fig:SYM-H}b).

The SYM-H index analysis for the day 26 December, 2004 showed a maximum positive value of $8~\mathrm{nT}$ during the period of $03:26$ UT and $03:44$ UT and again at $14:53$ UT and the minimum negative value of $-5~\mathrm{nT}$ occurred at $17:00$ UT (see Figure~\ref{fig:SYM-H}a).
In first level of decomposition, $(d^j)^2$($j=1$) showed two maximum of wavelets coefficients at  $08:10$ UT and at  $10:38$ UT.
In second level, $(d^j)^2$($j=2$) presented the three highest values at $07:28$ UT, $08:20$ UT and $11:44$ UT.
Only one maximum of wavelets coefficients $(d^j)^2$($j=3,4$) are shown in the third  and fourth levels at $07:28$ UT and $01:36$ UT, respectively.
Geomagnetic activity is classified by ``size'' (the amplitude of the variation in the magnetic records)  and usually described by the variation of indices to distinguish between a quiet and an active day (occurrence of storm or substorm).
The characteristic signature of a magnetic storm is a depression in the horizontal component of the Earth's magnetic field H at middle to low latitude ground stations. 
The index most used in low latitudes is the Dst index, similar to the SYM-H index.
However SYM-H is a minutely index, while Dst is a hourly index.
This is the reason that the wavelet treatment were done on SYM-H dataset.  
They represent the variations of the H component due to changes of the ring current \citep{Gonzalezetal:1994}. 
Following the terminology of Sugiura and Chapman discussed by \Citet{Gonzalezetal:1994}, the 26 December, 2004 presented a minimum Dst value of $-22$ nT at $19:00$ UT which corresponds to a weak geomagnetic storm.
As showed on Figure~\ref{fig:SYM-H}a, this day did not presented abrupt variations and consequently, in all the decompositions levels, the wavelet coefficients were quite not big enough.

We also analyzed the SYM-H index of the day after the Sumatra-Andaman tsunami.
On the 27 December, 2004, the SYM-H presented the maximum positive value of $14$ nT at $14:55~\mathrm{nT}$ and the minimum negative value of $-13~\mathrm{nT}$ at $21:36$ UT and at $23:44$ UT.
The Dst index showed minimum negative value of $-23~\mathrm{nT}$ between $22:00$ UT and $24:00$ UT which corresponds to a weak geomagnetic storm.
\Citet{MendesMag2005} discussed that when the magnetosphere is under quiet conditions or the analyzed function is smooth, the wavelets coefficients show very small amplitudes compare to when a geomagnetic storm is under development or abrupt functions.
The first and second level of decomposition showed the maximum of wavelets coefficients $(d^j)^2$($j=1,2$) at $09:16$ UT and at $09:16$ UT, respectively.
Both the third and fourth level showed the maximum of wavelets coefficients $(d^j)^2$($j=3,4$) at $07:12$ UT (see Figure~\ref{fig:SYM-H}b).
The wavelet coefficients also were quite not big enough.

As the results are referred to the analysis of a pos-processed magnetic dataset (SYM-H), the wavelet coefficients were expected to be quite not big enough.
The processing of the magnetic records in order to create a geomagnetic index acts as a filtering process that eliminates discontinuities.
So, a result of non identifying tsunami features was already expected. 
In the next steps we are going to deal directly with the records (magnetograms) of the selected magnetic stations.

\subsection{26-December-2004 tsunami event}

\begin{figure}[hbt]
\noindent
\centering
\begin{tabular}{cc}
(a) LRM, $26^{th}$ December, $2004$  & (b) EYR, $26^{th}$ December, $2004$ \\
\includegraphics[height=6.5cm,width=6.5cm]{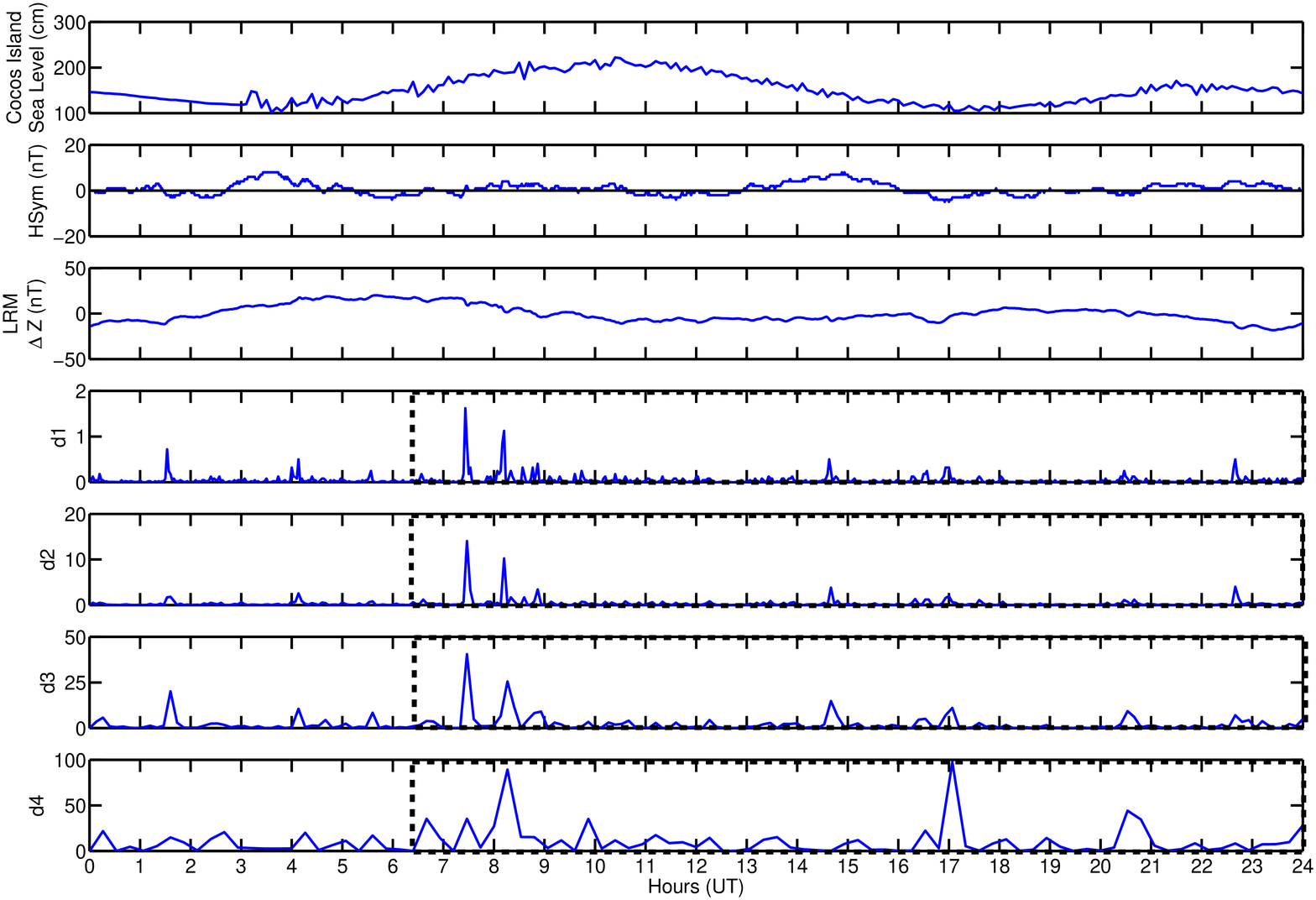}&
\includegraphics[height=6.5cm,width=6.5cm]{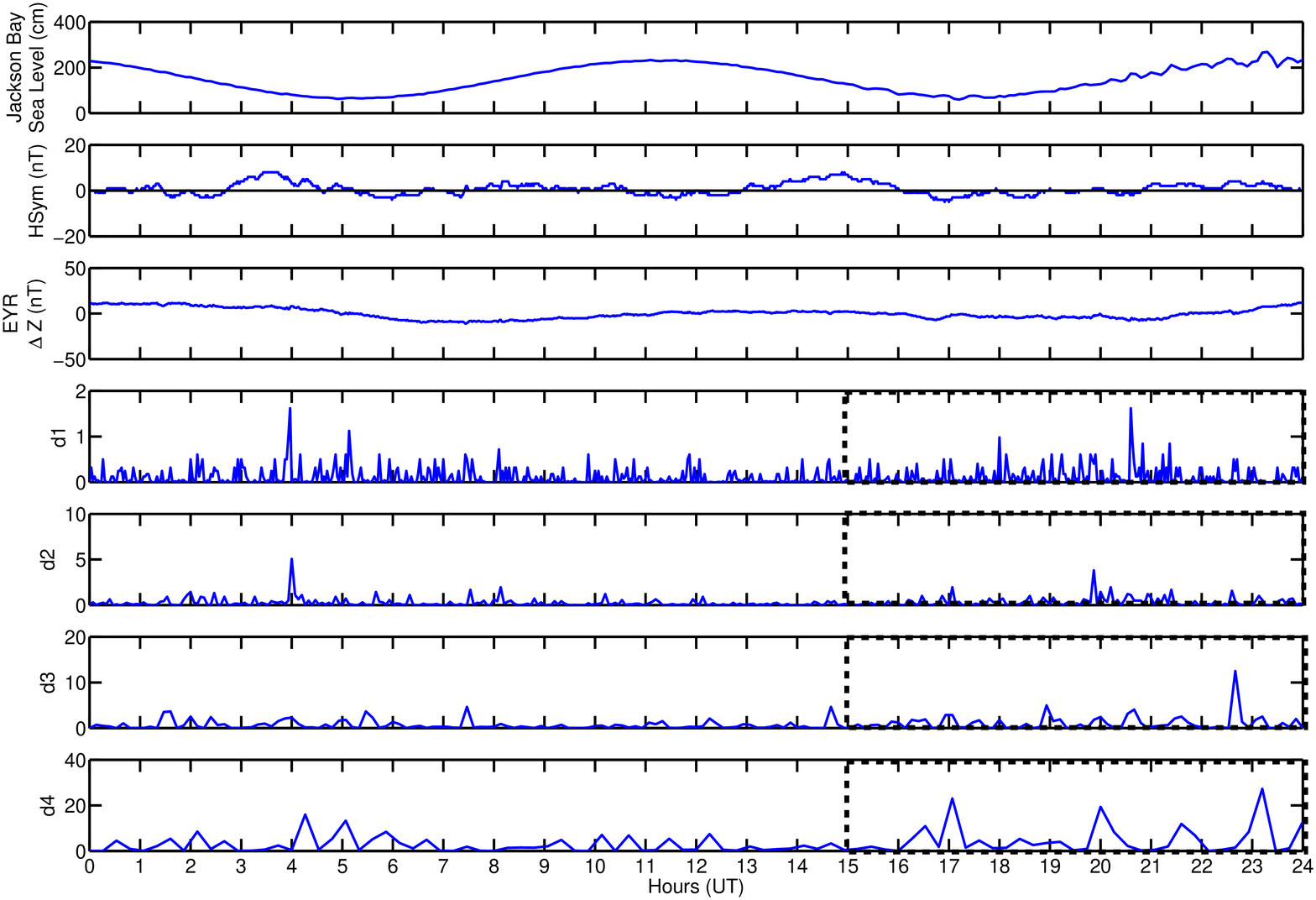}\\
(c) PPT, $26^{th}$ December, $2004$  & (d) HUA, $27^{th}$ December, $2004$ \\
\includegraphics[height=6.5cm,width=6.5cm]{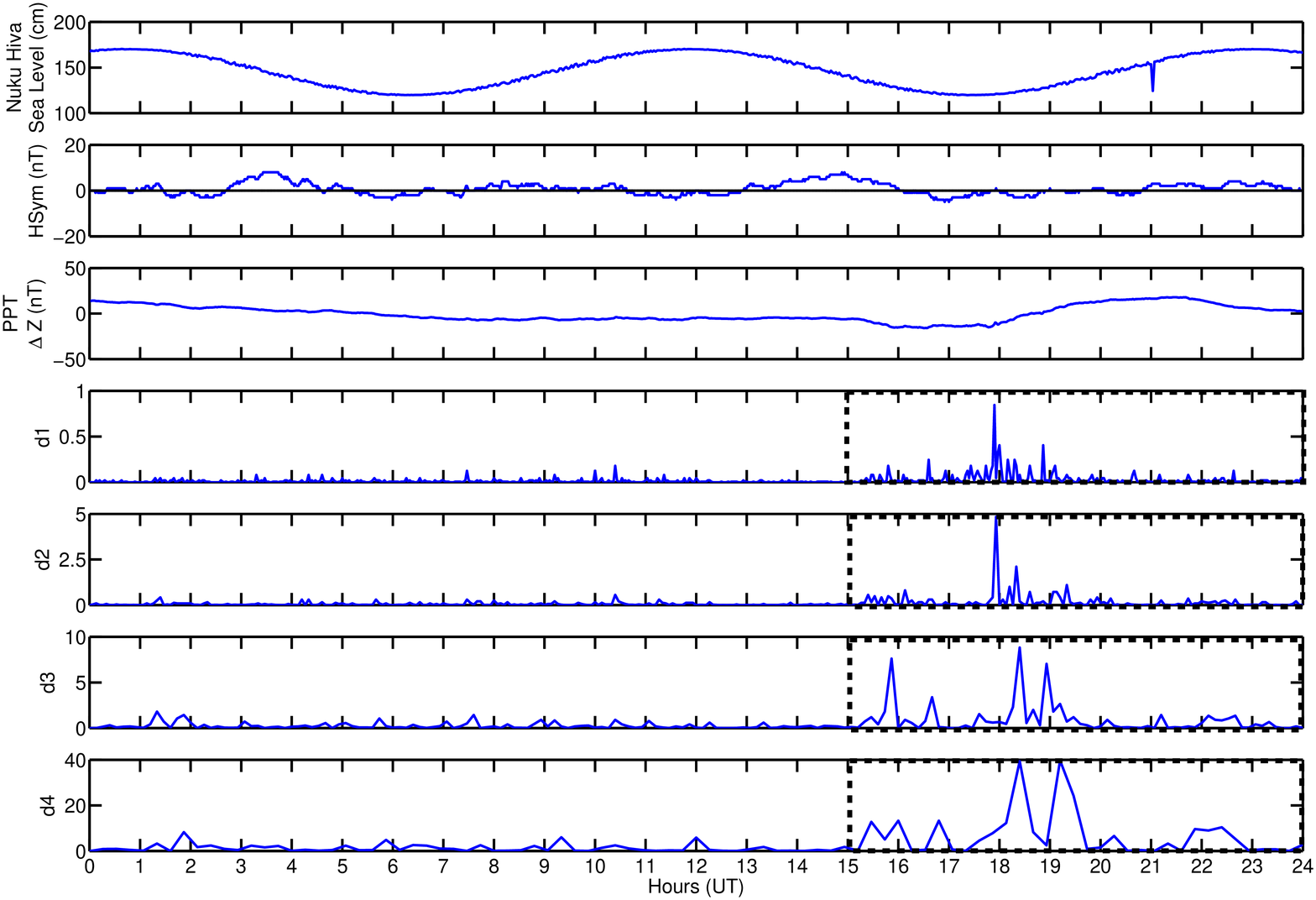} &
\includegraphics[height=6.5cm,width=6.5cm]{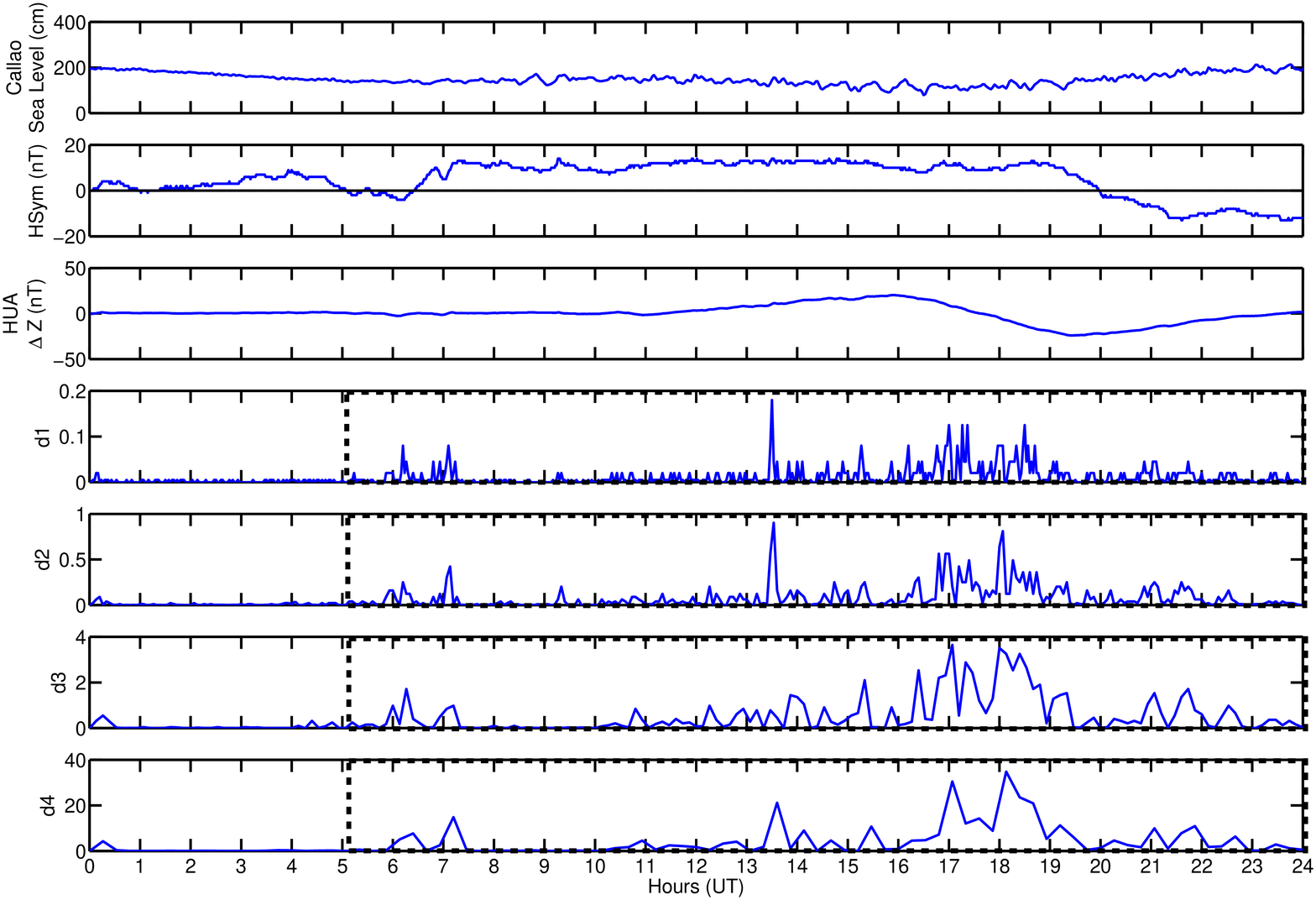}\\
\end{tabular}
\caption{Magnetograms and the wavelet coefficients $(d^j)^2$ for $j=1,2,3,4$ (a-d) for LRM, EYR, PPT and HUA, respectively.}
\label{fig:26DecDecomposition}
\end{figure}

In Figure~\ref{fig:26DecDecomposition}, panels from (a) to (d) correspond to magnetic stations of LRM, EYR, PPT and HUA, respectively.
Each panel, from top to bottom, displays the corresponding tide gauge measurement (sea level), 
the SYM-H index (both have been included for comparison purposes), the corresponding magnetogram (Z component), and
the wavelet signatures established by the square wavelet coefficients for the four first decomposition levels, $(d^j)^2$ for $j=1,2,3,4$.
The dashed rectangle is related to the period of occurrence of the tsunami identified by the methodology presented in this work.

The initial tsunami wave at the Cocos Island atoll (Australia) arrived at $03:17$ UT and it was measured with amplitude of up to $0.33~\mathrm{m}$.
After the first wave arrived, following tsunami waves continued for hours.
As reported by the Department of Transport - Government of Western Australia (http://www.transport.wa.gov.au/imarine/19383.asp), the initial wave arrived on Exmouth (Lat. $-21.93^{\circ}$, Long. $114.13^{\circ}$), Australia around $06:30$ UT.
It is expected that the magnetic field induced by the arrival of the tsunami will be detected at LRM around $06:30$ UT and forward.
Unfortunately, only Cocos Island was the near city to the magnetic station of LRM that we were able to acquire the tide gauge measurement.

Figure~\ref{fig:26DecDecomposition}a shows for guiding purposes the sea level for Cocos Island.
This was the nearest station measurement provided by the \textit{National Oceanic and Atmospheric Administration} - NOOA (http://wcatwc.arh.noaa.gov/about/tsunamimain.php).
We detected a small amplitude coefficients at $04:08$ UT, $17:00$ UT and $20:28$ UT, in the four first decomposition levels.
These $(d^j)^2$($j=1,2,3,4$) could not been explained by disturbed geomagnetic activity.
Also $(d^j)^2$($j=1,2,3$) presented small amplitude coefficients at $14:38$ UT and $22:40$ UT.

In Jackson Bay, New Zealand, the initial wave arrived at $15:10$ UT with amplitudes up to $0.65$ m.
In order to verify the transients due to the tsunami we used the EYR magnetogram data (see Figure~\ref{fig:26DecDecomposition}b).
For $(d^j)^2$ ($j=1$), the highest wavelet amplitudes were detected at $03:58$ UT, $05:08$ UT, $18:00$ UT, $20:36$ UT and $21:22$ UT.
We are only detecting the highest amplitudes that do not correspond to the wavelet coefficients signature detected by the previously SYM-H index analysis.
The $(d^j)^2$ ($j=2$) presented the highest amplitude coefficients at $03:58$ UT and $19:52$ UT.
The maximum amplitude $(d^j)^2$ ($j=3$) value occurred at $22:40$ UT.
For $(d^j)^2$ ($j=4$), the highest wavelet amplitudes occurred at $04:16$ UT, $05:04$ UT, $17:04$ UT, $20:00$ UT, $21:36$ UT and $23:12$ UT.

Figure~\ref{fig:26DecDecomposition}c shows the wavelet decompositions for the PPT, French Polynesia.
As reported by NOAA, the initial wave arrival could not be determined exactly.
For guiding purposes, we presented the sea level variations at Nuku Hiva, French Polynesia.
At all the four decomposition levels, we detected perceptible $(d^j)^2$($j=1,2,3,4$) amplitudes between $15:00$ UT and $20:00$ UT that might be due to the tsunami propagation.

On 27 December, 2004, the tsunami waves that propagated throughout the Pacific Ocean arrived at the coast of Peru (see Figure~\ref{fig:26DecDecomposition}c).
At Callao, Peru, the initial tsunami wave arrived at $05:20$ UT with maximum wave height of $0.65$ m.
The $(d^j)^2$ for $j=1,2,3,4$ decomposition levels showed what might be tsunami transients between $13:00$ UT and $19:00$ UT.

Thus from the methodology applied, the transients in the magnetogram, that are not related to geomagnetic storms, are identified by the wavelet coefficients.
Those transients are not related to the common geomagnetic disturbance causes, but some remarkable coefficients are related to the ocean stream.
The dashed rectangles in each panel of the figure indicate the identification of the transients associated to the tsunami magnetic effects.

\subsection{27-February-2010 tsunami event}

\begin{figure}[hbt]
\noindent
\centering
\begin{tabular}{cc}
(a) HUA, $27^{th}$ February, $2010$  & (b) IPM, $27^{th}$ February, $2010$ \\
\includegraphics[height=6.5cm,width=6.5cm]{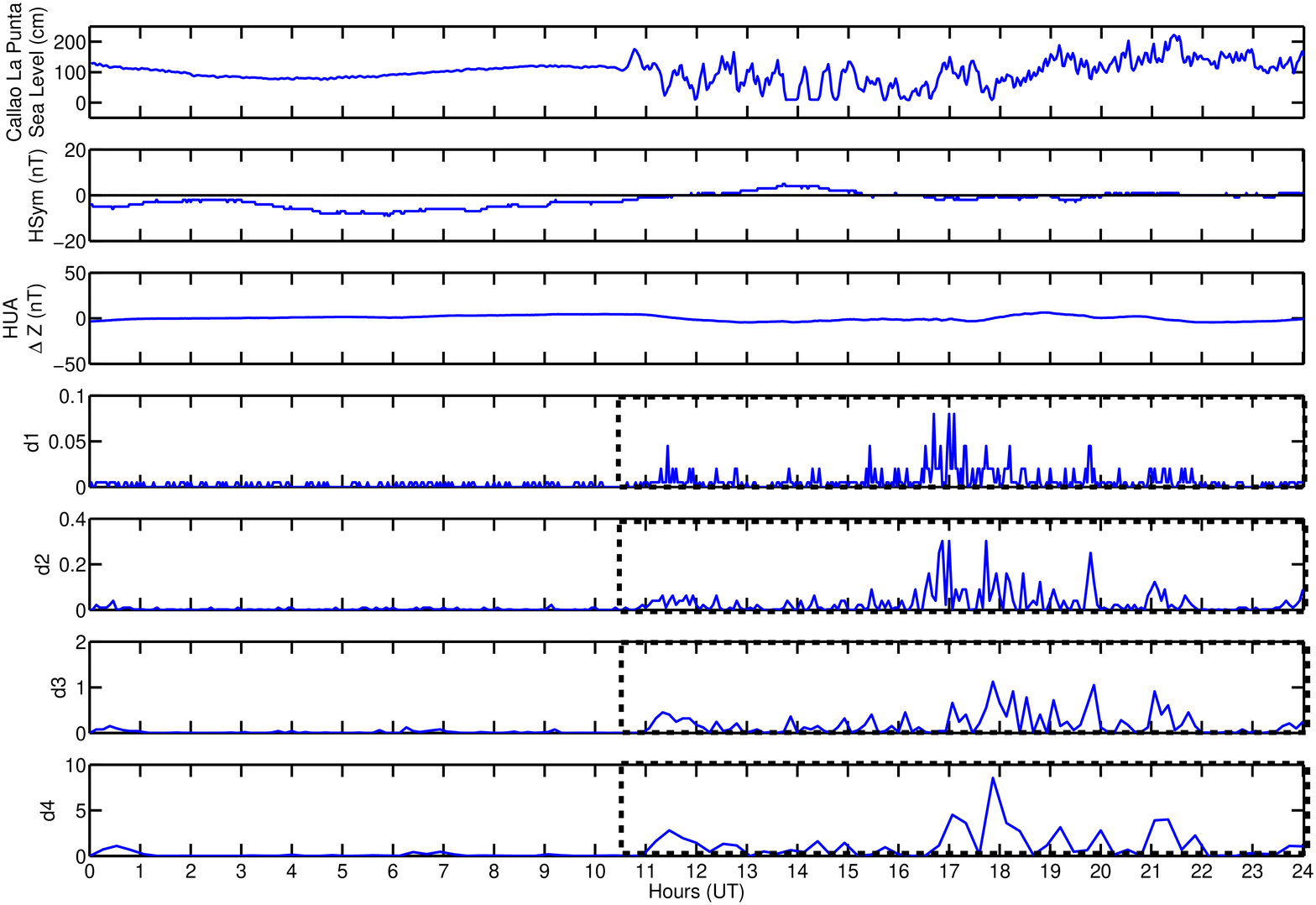}&
\includegraphics[height=6.5cm,width=6.5cm]{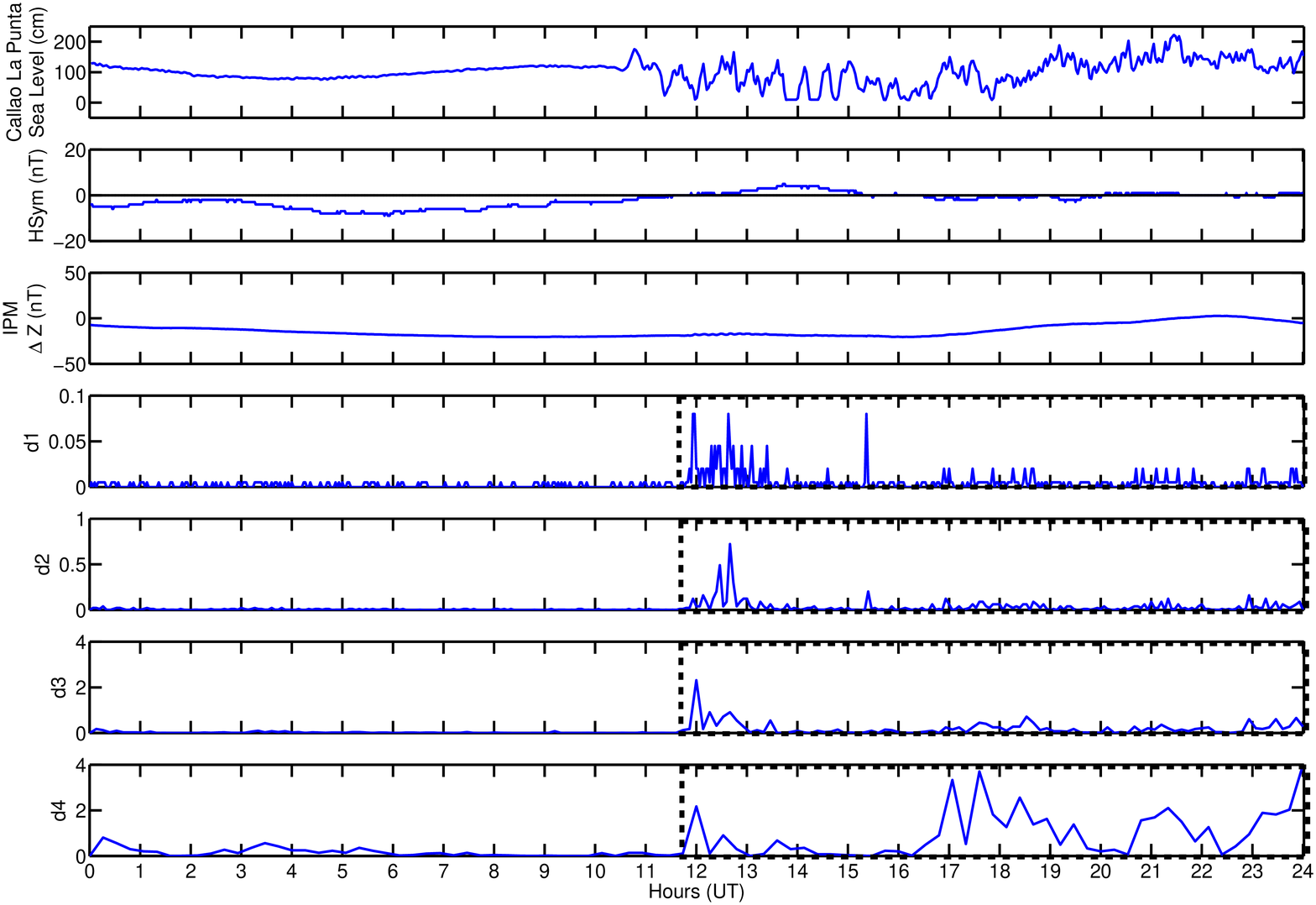}\\
(c) PPT, $27^{th}$ February, $2010$ & (d) TEO, $27^{th}$ February, $2010$ \\
\includegraphics[height=6.5cm,width=6.5cm]{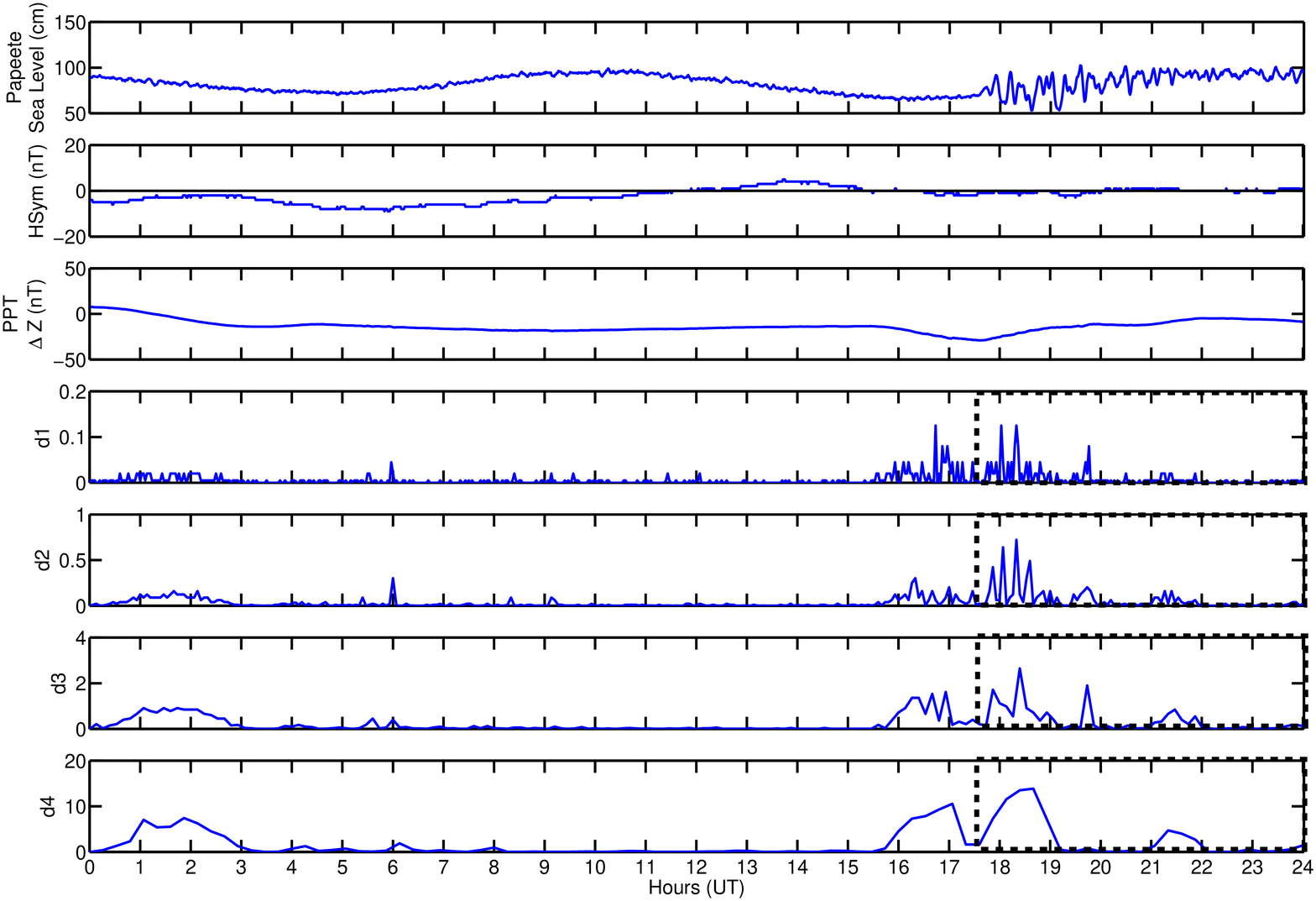} &
\includegraphics[height=6.5cm,width=6.5cm]{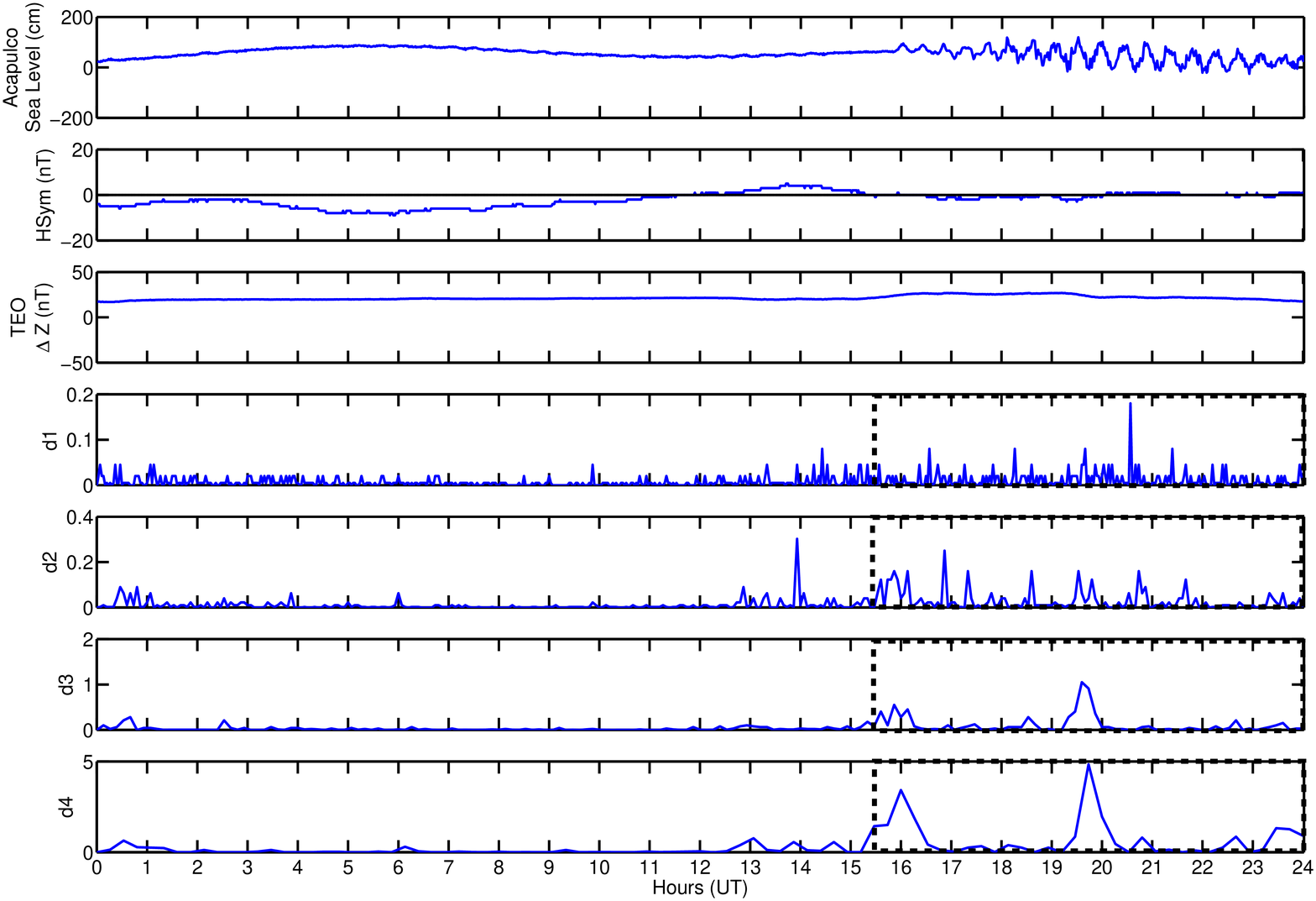}\\
\end{tabular}
\caption{Magnetograms and the wavelet coefficients $(d^j)^2$ for $j=1,2,3,4$ (a-d) for HUA, IPM, PPT and TEO, respectively.}
\label{fig:27FebDecomposition}
\end{figure}

In the region of Callao La Punta, Peru, the observed and computed tsunami time arrival were coincident, both at $10:34$ UT with amplitude of up to $0.69$ m.
We used the sea level measured at this region as guiding line to the arrival tsunami at HUA and IPM (see Fig~\ref{fig:27FebDecomposition}a-b).
Fig~\ref{fig:27FebDecomposition} is very similar to Figure~\ref{fig:26DecDecomposition}, also shows the wavelet signatures for the four first decomposition levels ($(d^j)^2$ where $j=1,2,3,4$).
Panels from (a) to (d) correspond to magnetic stations of HUA, IPM, PPT and TEO, respectively.
In each panel, from top to bottom are displayed the corresponding tide gauge measurement, the SYM-H index, the corresponding magnetogram (Z component), and the $(d^j)^2$ for $j=1,2,3,4$ decomposition levels.
On geomagnetic conditions, the 27 February, 2010 corresponded to a very quite day.
The Dst index presented the minimum of $-2~\mathrm{nT}$ and the maximum of $4~\mathrm{nT}$ with very smooth variations.
The SYM-H also presented very smooth variations with the minimum of $-9~\mathrm{nT}$ and the maximum of $5~\mathrm{nT}$.

In Fig~\ref{fig:27FebDecomposition}a, the $(d^j)^2$ for $j=1,2,3,4$ decomposition levels presented quite similar behaviors.
Also, they showed a main structure of coefficients which are preceded and followed by a sequence of small structures after $11:00$ UT.
The highest wavelet  coefficient amplitudes occurred at $11:26$ UT, $15:26$ UT, between  $16:42$ UT and $18:12$ UT, $19:48$ UT and between $21:00$ UT and $21:48$ UT.

The NOAA Pacific Warning Center predicted the tsunami time arrival at $12:05$ UT for the Easter Island, IPM.
In Fig~\ref{fig:27FebDecomposition}b, the first decomposition level presented a main structure of coefficients between $11:58$ UT and $13:24$ UT and a secondary pike at $15:36$ UT, also presented a sequence of small structures between $16:54$ UT and $24:00$ UT.
$(d^j)^2$($j=2$) showed the highest coefficients at $12:28$ UT and $12:40$ UT and $(d^j)^2$($j=3$) at $12:00$ UT.
$(d^j)^2$($j=2,3$) presented less structured features than $(d^j)^2$($j=1,4$).
In $(d^j)^2$($j=4$) was possible to notice two peaks, one at $12:00$ UT and the other at $12:32$ UT, followed by a main structure of coefficients between $17:04$ UT and $19:28$ UT and a secondary structure between $20:48$ UT and $21:06$ UT.

Fortunately, for PPT, we were able to get the tide gauge measurements at the same location (see Fig~\ref{fig:27FebDecomposition}c).
The observed tsunami initial arrival time was at $17:33$ UT and the computed time was at $17:47$ UT with amplitude up to $0.22~\mathrm{m}$.
The $(d^j)^2$($j=1,2,3,4$) presented surprisingly similar wavelet signatures.
In Fig~\ref{fig:27FebDecomposition}c, the main wavelet coefficient structures are restricted between $15:36$ UT and $21:52$ UT.
The tsunami-induced electromagnetic field effects could be detected about two hours in advance.

We also selected a further located station TEO in Mexico with the purpose of analyzing if we could detect any geomagnetic tsunami effects.
For guiding purposes, we used the tide gauge measurements at Acapulco.
The observed tsunami initial arrival time was at $15:47$ UT with amplitude up to $0.66~\mathrm{m}$.
In Fig~\ref{fig:27FebDecomposition}d, the $(d^j)^2$($j=1$) showed an even more structured signature than $(d^j)^2$($j=2,3,4$).
The highest wavelet amplitudes coefficients for $(d^j)^2$($j=1$)  was at $14:26$ UT, $16:34$ UT, $18:16$ UT, $19:40$ UT, $20:34$ UT and $21:24$ UT.
The $(d^j)^2$($j=2,3,4$) presented a quite similar signature but less structured features than $(d^j)^2$($j=1$).
It was possible notice the presence of highest wavelet amplitudes coefficients after $12:52$ UT for  $(d^j)^2$($j=2,3,4$) that could be related to the tsunami effects.

\subsection{11-March-2011 tsunami event}

On $11$ March, $2011$, occurred a geomagnetic disturbance which corresponded to a moderate storm with minimum Dst $=-82~\mathrm{nT}$ at $06:00$ UT and a  second energy injection at $18:00$ UT with minimum Dst $=-67~\mathrm{nT}$ at $22:00$ UT. 
So, the abrupt variations of the vertical component of the geomagnetic field due to the  development of this moderate storm will be more emphasized by the highest amplitudes of the wavelet coefficients.

\begin{figure}[hbt]
\noindent
\centering
\begin{tabular}{c}
$11^{th}$ March, $2011$ \\
\includegraphics[height=6.5cm,width=6.5cm]{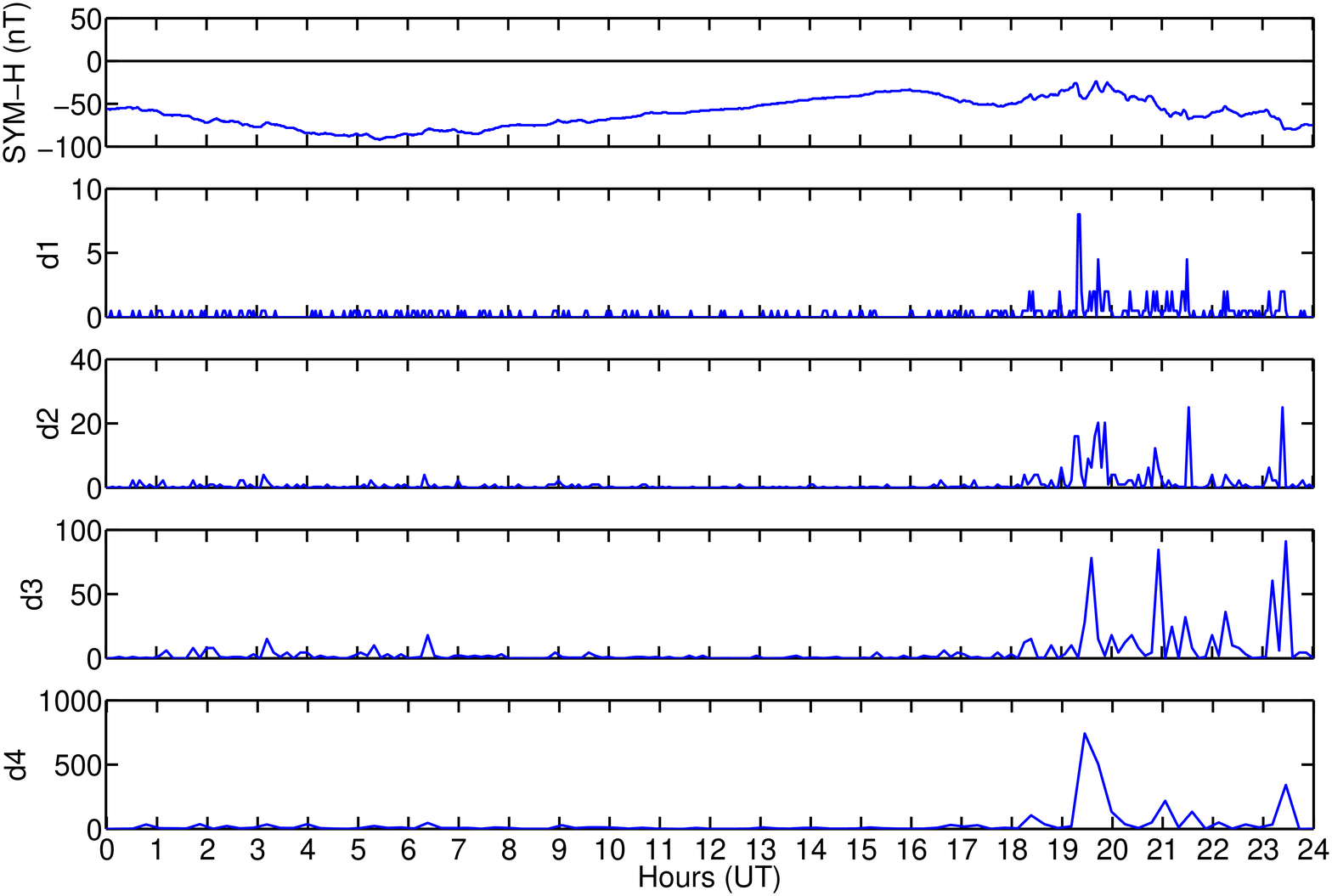}\\
\end{tabular}
\caption{Graphics of SYM-H with the corresponding wavelet coefficients $(d^j)^2$ for $j=1,2,3,4$. 
The panel corresponds to a japanese tsumani occurrence at 2011.}
\label{fig:MarchSYM-H}
\end{figure}

In Figure~\ref{fig:MarchSYM-H}, the highest coefficients of all the four levels detected in the SYM-H are restricted to the development of the storm following the second energy injection at $18:00$ UT, specially during the main phase of the geomagnetic storm between  $18:00$ UT and  $24:00$ UT presented the highest wavelet coefficients amplitudes which was usually observed by \cite{Domingues2005}.
All the decomposition levels show quite similar behaviors, showing the highest wavelet coefficients $(d^j)^2$($j=1,2,3,4$) soon after the recovery time of the first storm and associated with the fluctuations of the SYM-H index as the second magnetic storm came to the minimum value.

\begin{figure}[hbt]
\noindent
\centering
\begin{tabular}{cc}
(a) MMB, $11^{th}$ March, $2011$  & (b) KNY, $11^{th}$ March, $2011$  \\
\includegraphics[height=6.5cm,width=6.5cm]{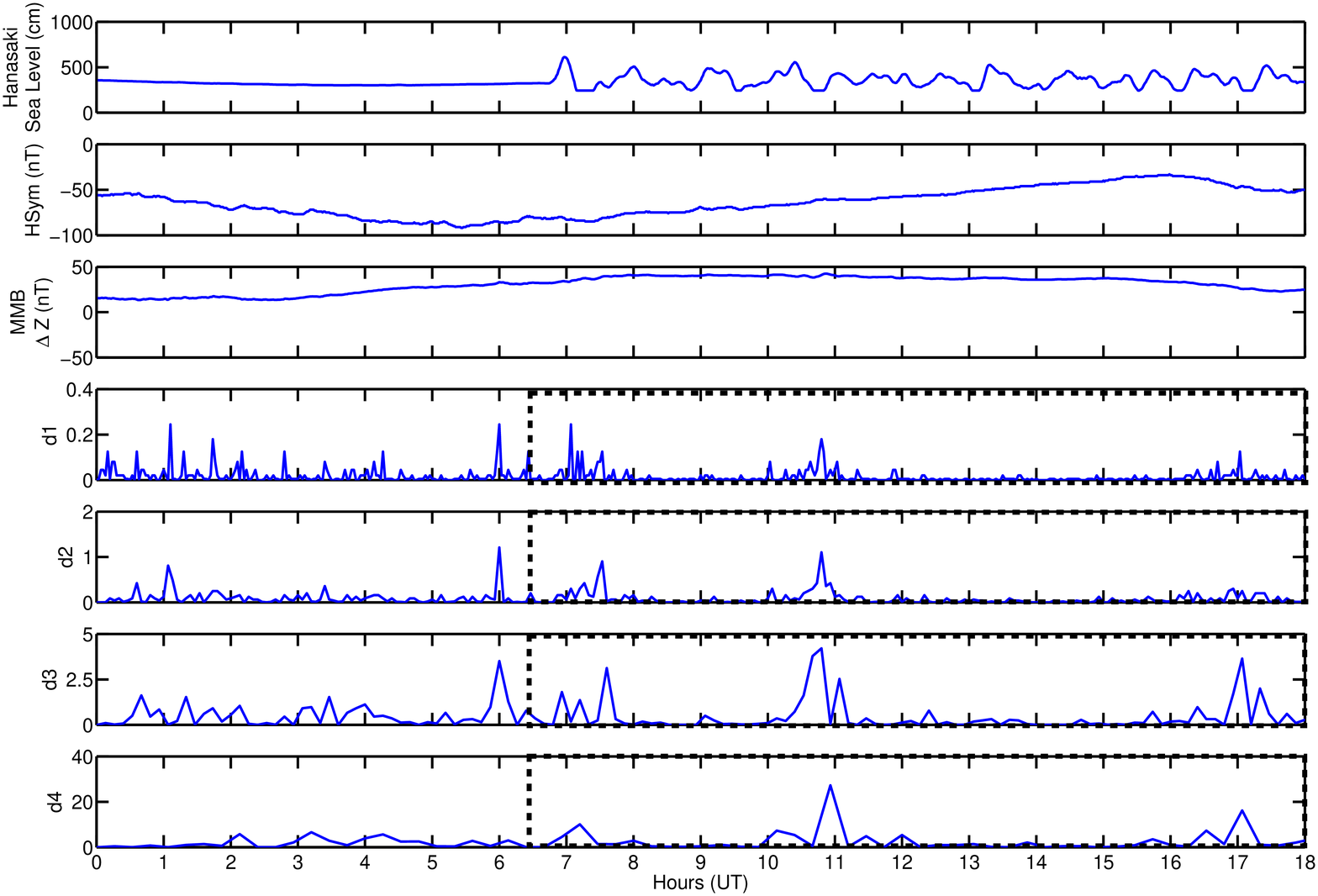}&
\includegraphics[height=6.5cm,width=6.5cm]{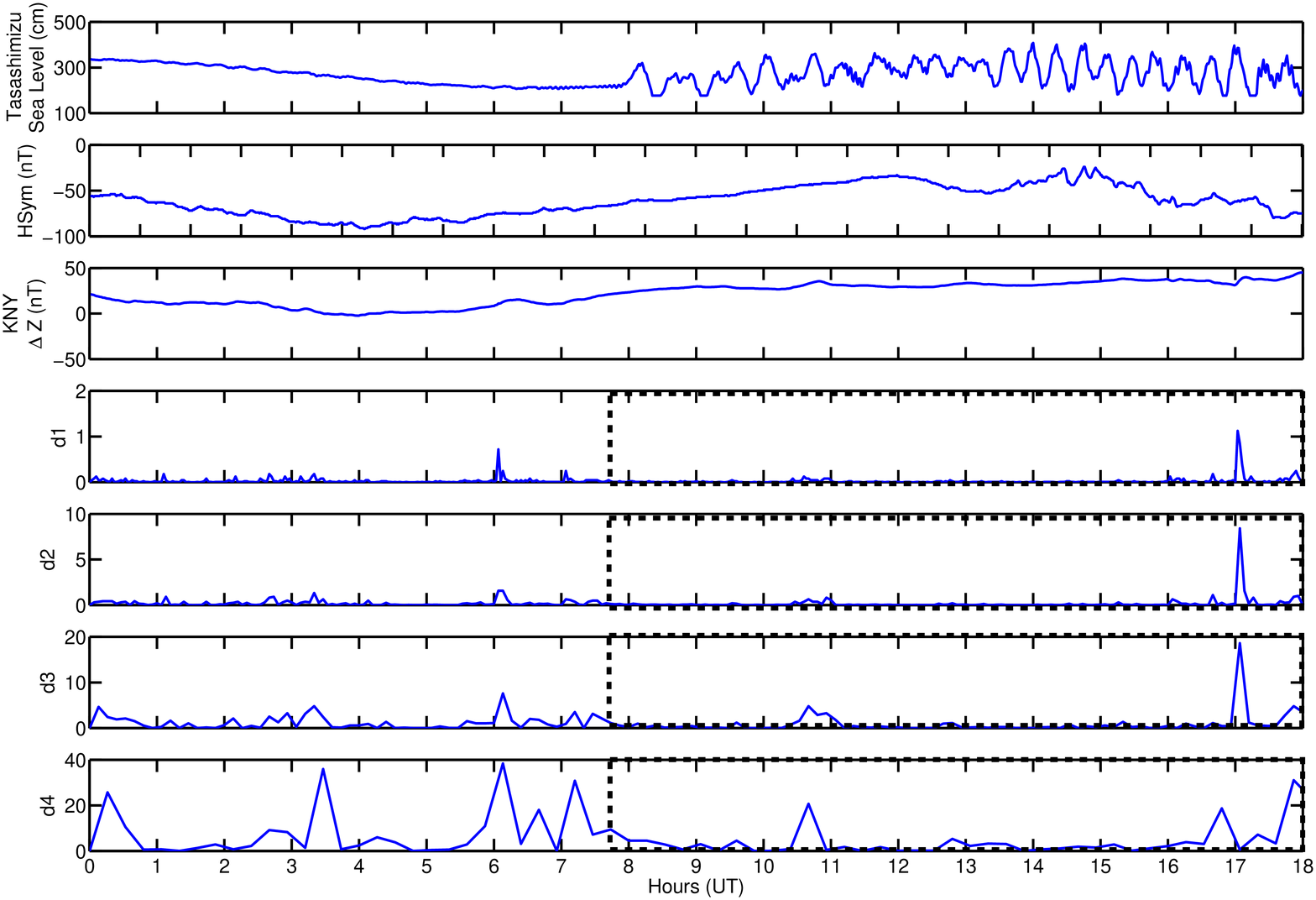}\\
(c) GUA, $11^{th}$ March, $2011$  & (d) CTA, $11^{th}$ March, $2011$ \\
\includegraphics[height=6.5cm,width=6.5cm]{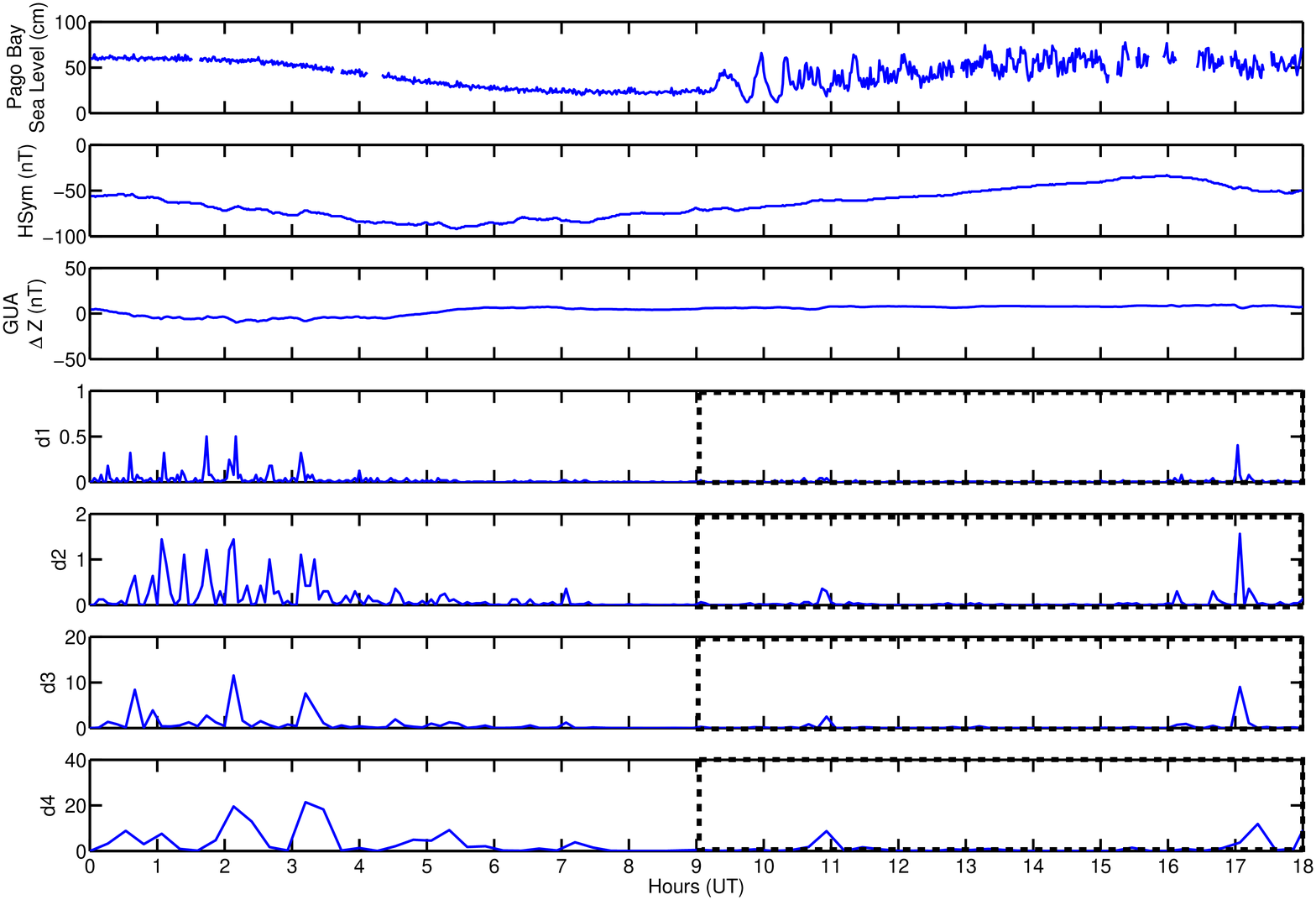} &
\includegraphics[height=6.5cm,width=6.5cm]{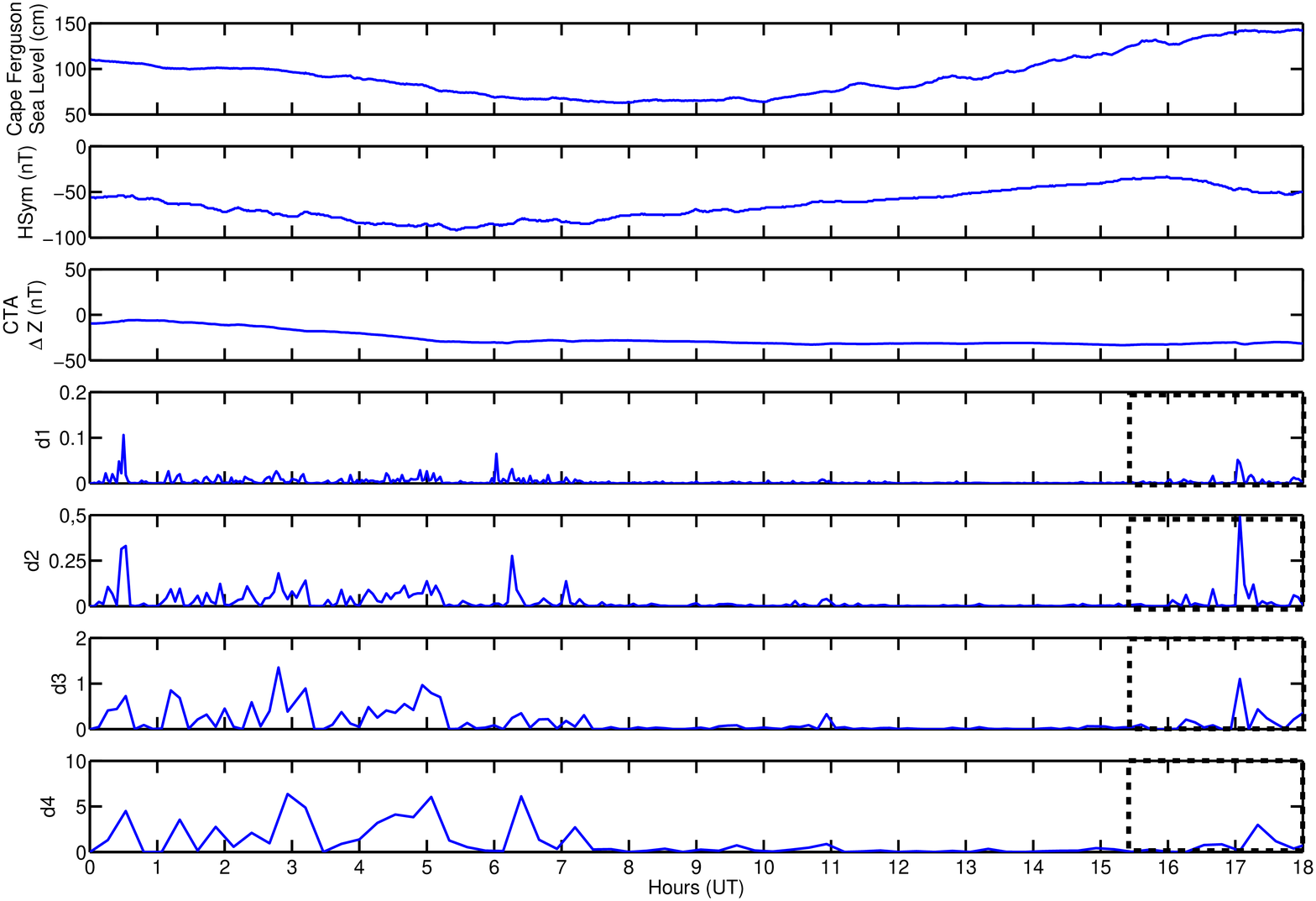}\\
\end{tabular}
\caption{Magnetograms and the wavelet coefficients $(d^j)^2$ for $j=1,2,3,4$ (a-d) for MMB, KNY, GUA and CTA, respectively.}
\label{fig:11MarDecomposition}
\end{figure}

Figure~\ref{fig:11MarDecomposition} is very similar to Figure~\ref{fig:26DecDecomposition} and Fig~\ref{fig:27FebDecomposition}.
It also shows the wavelet signatures for the four first decomposition levels, $(d^j)^2$ for $j=1,2,3,4$.
Panels from (a) to (d) correspond to magnetic stations of HUA, IPM, PPT and TEO, respectively.
In each panel, from top to bottom are displayed the corresponding tide gauge measurement, the SYM-H index (both have been included for comparison purposes only), the corresponding magnetogram (Z component), and the $(d^j)^2$ for $j=1,2,3,4$ decomposition levels.
We decided to exclude the period between $18:00$ UT and  $24:00$ UT, as shown in Figure~\ref{fig:MarchSYM-H}, because the wavelet coefficients in this period were related to the abrupt variations caused by the second energy injection in the geomagnetic storm.

In the city of Hanasaki, near to the magnetic station of MMB, the initial phase of the tsunami started at $06:38$ UT and reached the maximum height ($0.74$ m) at $06:57$ UT.
We used the tide gauge measurements at Hanasaki in Figure~\ref{fig:11MarDecomposition}a for guiding purposes.
The $(d^j)^2$($j=1,2,3,4$) showed a very structured signature.
Also, $(d^j)^2$($j=1,2,3$) showed a main structure of wavelet coefficients that could be noticed between $06:00$ UT and $08:00$ UT, a secondary structure between $10:00$ UT and $11:20$ UT and a smaller structure between $16:00$ UT and $18:00$ UT.
The $(d^j)^2$($j=4$) showed less structured features than $(d^j)^2$($j=1,2,3$) and presented the highest amplitudes at $07:12$ UT, $10:56$ UT and $17:04$ UT.
We can observe at the first decomposition level, a presence of wavelet coefficients at $06:26$ UT which corresponds to the tsunami initial phase.
At $06:57$ UT, when tsunami reached the maximum height, we also notice the presence of an increase in the coefficient amplitudes with the maximum at $07:04$ UT higher than the coefficient at the initial phase.
Meanwhile, $(d^j)^2$($j=2,3,4$) associated to the tsunami presented more spread on time and did not present any remarkable amplitudes corresponding to its initial phase and maximum height.

Near to the KNY station, in the city of Tosashimizu, the tsunami started at $07:51$ UT and the maximum height ($1.28$m) was measured at $16:58$ UT (see Figure~\ref{fig:11MarDecomposition}b).
The wavelet signatures of  $(d^j)^2$($j=1,2,3,4$) show very similar structures.
The $(d^j)^2$($j=1,2$) showed the highest wavelet amplitudes around $06:04$ UT and $17:02$ UT.
And the $(d^j)^2$($j=3,4$) showed the highest wavelet amplitudes around $06:08$ UT, $06:40$ UT, $07:44$ UT, $10:40$ UT, $16:48$ UT and $17:52$ UT.
This japanese magnetic station showed surprisingly similar signatures compared to  Figure~\ref{fig:11MarDecomposition}a.
We also observe, in third and fourth decomposition levels, the presence of wavelet coefficients $(d^j)^2$($j=3,4$) around the tsunami initial phase and maximum height.

In Figure~\ref{fig:11MarDecomposition}c, the observed time of the tsunami's arrival at Pago Bay, Guam around $09:16$ UT.
The tsunami maximum height occurred at $09:58$ UT and its amplitude was up to $0.54~\mathrm{m}$.
GUA magnetic station did not show any coefficient $(d^j)^2$($j=1,2,3,4$) that represented significant contribution due to the tsunami.
At the same time, we observe the presence of small coefficients $(d^j)^2$($j=1,2,3,4$) but still noticeable between $10:40$ UT and $11:28$ UT.

The predicted time of the tsunami's arrival by NOAA at Cairns (Lat. $-16.57^{\circ}$ and Long. $145.75^{\circ}$), Australia, would be at $15:35$ UT 
Unfortunately, at Cape Ferguson, it was not possible to determine the exactly tsunami time arrival (see Figure~\ref{fig:11MarDecomposition}d).
The tide gauge measurements registered the tsunami maximum height at $21:00$ UT with amplitudes up to $0.11~\mathrm{m}$.
The CTA magnetic station did not show any remarkable wavelet coefficient amplitudes $(d^j)^2$($j=1,2,3,4$) due to the tsunami.

In summary, the wavelet coefficient of the first three levels of decomposition showed a good time localization of the initial phase and the maximum height of the Japanese tsunami at $11$th of March, $2011$ and they, also, were locally associated with the geomagnetic variations present in the signal due to the secondary magnetic fields induced by the movement of electrically conducting sea-water through the geomagnetic field.
The red vertical lines in each panel of the figure indicate the identification of the first transient associated to the tsunami magnetic effects.

Thus from the several events analyzed the wavelet method for identification of the tsunami magnetic effects could be validated, characterizing an useful tool for this kind of study in order to deal with very weak but significant amplitude of wavelet coefficients.

\section{Conclusions}
\label{Conclusions}

Dealing with the geomagnetic data analyzed by discrete wavelet transform using Haar wavelet, this work aimed to implement a methodology for the direct analysis of magnetograms and to verify the occurrence of tsunami effects on Z-component of the geomagnetic field.
The carried out analysis can be summarized as follows:

\begin{itemize}
\item The wavelet technique is useful to ``zoom in'' the localized behavior of the geomagnetic variations induced by the movement of electrically conducting sea-water through the geomagnetic field; \textit{i.\,e.}, the identification of magnetic transients related to the tsunamis. As well, from the analysis of the magnetogram data, it was able to localize the initial phase and maximum height of the tsunamis in some cases.
\item The discrete wavelet transform (Daubechies - db1) provides a very good local characterization of the signal and may be the most convenient to represent times series with abrupt variations or steps, \textit{i.\,e.}, very small and localized variations as the discontinuities in the Z-component due to tsunamis.
\item The discrete wavelet transform is an alternative way to analyze the global influence of the tsunamis on the geomagnetic field and it could be used as a sophisticated tool to predict the tsunami arrival, as an example of the Chilean tsunami, the DWT detected transients about two hours in advance at Papeete.
\end{itemize}


The results obtained are encouraging even with a few of events.
The present study dealt with only three tsunami events and few magnetic stations.
A future study using more events and more stations should be very good in order to do a more complete analysis of features.
The first interpretation of the results suggests that discrete wavelet transform can be used to characterize the tsumanis effects on the geomagnetic field.

\section{Acknowledgments}

V. Klausner wishes to thanks CAPES for the financial support of her PhD.
A. R. R. P. thanks CNPq for a research fellowship.
This work was supported by CNPq (grants 309017/2007-6, 486165/2006-0, 308680/2007-3, 478707/2003, 477819/2003-6, 382465/01-6), FAPESP (grants
2007/07723-7) and CAPES (grants 86/2010-29). 
Also, the authors would like to thank the INTERMAGNET programme for the datasets used in this work.

\bibliographystyle{abbrv}

\end{document}